\documentclass[a4paper]{jpconf}
\usepackage[dvipsnames]{xcolor}
\usepackage{colortbl}
\definecolor{red}{rgb}{0.77255,0.05490,0.12157}%
\definecolor{lightred}{rgb}{0.60784,0.21960,0.21960}%
\definecolor{green}{rgb}{0.4627,0.72549,0.0}%
\definecolor{blue}{rgb}{0.00000,0.21961,0.50980}%
\definecolor{white}{rgb}{1.00000,1.00000,1.00000}%
\definecolor{violet}{rgb}{0.40784,0.13333,0.54510}%
\definecolor{orange}{rgb}{1.00000,0.49804,0.00000}%
\definecolor{brown}{rgb}{0.54510,0.27059,0.07451}%
\definecolor{grey100}{rgb}{0.3921,0.3921,0.3921}%
\definecolor{grey150}{rgb}{0.5882,0.5882,0.5882}%
\definecolor{grey200}{rgb}{0.7843,0.7843,0.7843}%
\usepackage{graphicx}
\usepackage{amsmath}
\DeclareMathOperator{\sinc}{sinc}
\usepackage{subfig}
\usepackage{tikz}
\usepackage{tikzscale}
\usepackage{pgfplots}
\usepackage{grffile}
\pgfplotsset{compat=newest}
\usetikzlibrary{plotmarks}
\usetikzlibrary{arrows.meta}
\usepgfplotslibrary{patchplots}
\usetikzlibrary{patterns}
\DeclareRobustCommand{\legendsquare}[1]{%
	\tikz[baseline=(a.south)]{\node[#1, draw, line width=.1ex, inner sep=.7ex, outer sep=0] (a) {};}%
}

\usepackage{placeins}

\begin{document}
\title{Topology comparison of superconducting AC machines in hybrid-electric aircraft}

\author{Matthias Corduan\textsuperscript{1}, Martin Boll\textsuperscript{2}, Roman Bause\textsuperscript{2}*, Marijn P. Oomen\textsuperscript{1}, Mykhaylo Filipenko\textsuperscript{2}, Mathias Noe\textsuperscript{3}}

\address{\textsuperscript{1} Siemens AG, Corporate Technology, Postbox 32 20, 91050 Erlangen, Germany}
\address{\textsuperscript{2} Siemens AG, eAircraft, Willy-Messerschmitt-Str. 1, 82024 Taufkirchen, Germany}
\address{\textsuperscript{3} Institute of Technical Physics, Karlsruhe Institute of Technology, 76344 Eggenstein-Leopoldshafen, Germany}
\address{*Present address: Max-Planck-Institute of Quantum Optics, 85741 Garching, Germany}

\ead{matthias.corduan@siemens.com}

\begin{abstract}
Electric machines with very power-to-weight ratios are inevitable for hybrid-electric aircraft applications. One potential technology that is very promising to achieve the required power-to-weight ratio for short-range aircraft, are superconductors used for high current densities in the stator or high magnetic fields in the rotor. In this paper, we present an in-depth analysis of the potential for fully and partially superconducting electric machines that is based on an analytical approach taking into account all relevant physical domains such as electromagnetics, superconducting properties, thermal behavior as well as structural mechanics. For the requirements of the motors in the NASA N3-X concept aircraft, we find that fully superconducting machines could achieve 3.5 times higher power-to-weight ratio than partially superconducting machines. Furthermore, our model can be used to calculate the relevant KPIs such as mass, efficiency and cryogenic cooling requirements for any other machine design.
\end{abstract}

\section{Introduction}

Stringent emission targets have been decided by the European Commission in the Flightpath 2050 vision, which demands a 75\% reduction in CO\textsubscript{2}, a 90\% reduction in NO\textsubscript{x} and 65\% lower perceived noise emissions, compared with a typical new aircraft in 2000 \cite{EuropaischeKommission.2011}. One possible way to achieve these goals is by replacing the conventional drive train system by a hybrid-electric one. Detailed studies on aircraft level propose that remarkable reductions in emissions can be achieved. For instance, with the N3-X concept aircraft - a turbo-electrically powered blended-wing body aircraft for roughly 300 PAX - could have less 70\% emission than a comparable conventionally powered design without any additional restriction on cruise speed or payload. However, the technological feasibility of this concept highly depends on the availability of very light-weight electric machines and components. For the N3-X electric machines with a power-to-weight ratio $PTW$ of at least $12.7\,\mathrm{kW\,kg^{-1}}$ are required \cite{Felder.2011}. \\
As the PTW of electric machines is proportional to airgap magnetic field and the stator current loading ($PTW \sim B_{\mathrm{gap}} \cdot A$), superconducting materials offer great potential having rotor fields larger than with permanent magnets and much higher current densities in the stator when compared to copper. Thus, many authors claim that combining both advantages could result in machines with a very high power-to-weight ratio. However, as the AC-losses in superconductors scale non-linear with the exposed magnetic field and electric frequency, several effects counteract each other and a detailed analysis on machine design level is necessary to determine the optimal trade-off.\\ 
In this paper, we outline such an analysis based on analytical models for most relevant physical domains. We compare the potential of a fully superconducting machine (i.e. with superconducting rotor and stator) and a partially superconducting machine (i.e. only with a superconducting rotor) concerning their most relevant KPIs, such as mass and efficiency for a given set of high-level requirements such as power and rotation speed. We present an analysis based on the motor requirements from the N3-X concept aircraft design. However, the model can easily be applied to any other set of high-level requirements. Further, the choice of a radial flux topology is motivated by symmetry considerations that make it easier to calculate the magnetic potential analytically. Nevertheless, it can be expected, that our general findings would be similar for an axial flux topology. \\
The paper is structured as follows: In Section 2 the details of the analytical calculation models are presented. In Section 3 the results for the N3-X requirements are presented followed by the conclusions in Section 4.

\section{Machine model}
\begin{figure}[!h]
	\centering       
	\def\svgwidth{0.48\textwidth}    
	\includegraphics{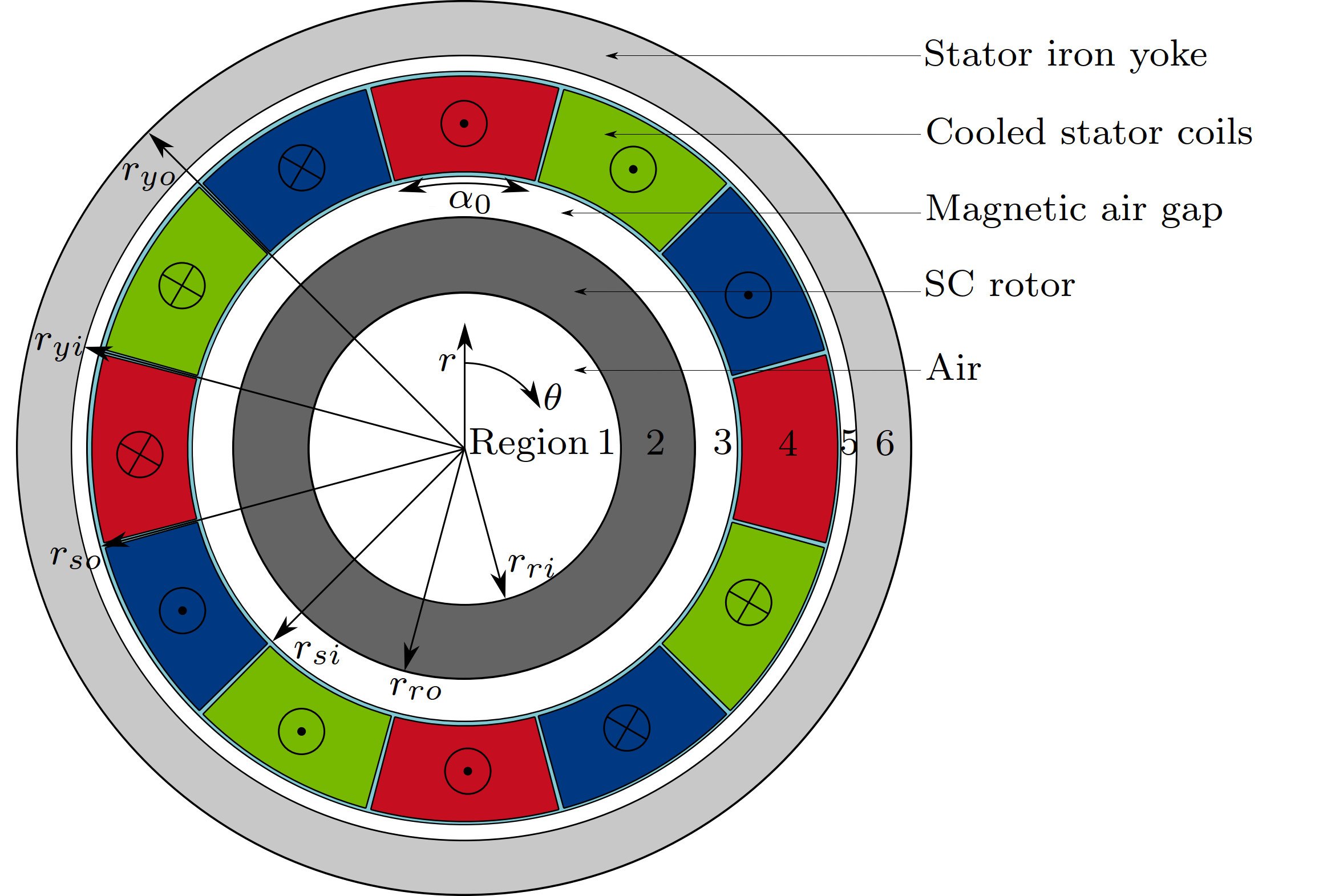}
	\caption{General scheme of a synchronous superconducting radial flux machine with a three-phase winding system marked \legendsquare{fill=red}, \legendsquare{fill=blue}, \legendsquare{fill=green}. Geometry parameters are the inner rotor radius $r_{ri}$, the outer rotor radius $r_{ro}$, the inner stator radius $r_{si}$, the outer stator radius $r_{so}$, the inner yoke radius $r_{yi}$, the outer yoke radius $r_{yo}$ and the duty cycle of the stator coils $\alpha_{0}$.}    
	\label{fig:rfm_model_cold_airgap} 
\end{figure}

\begin{figure}[!h]
	\centering       
	\def\svgwidth{0.47\textwidth}    
	\includegraphics{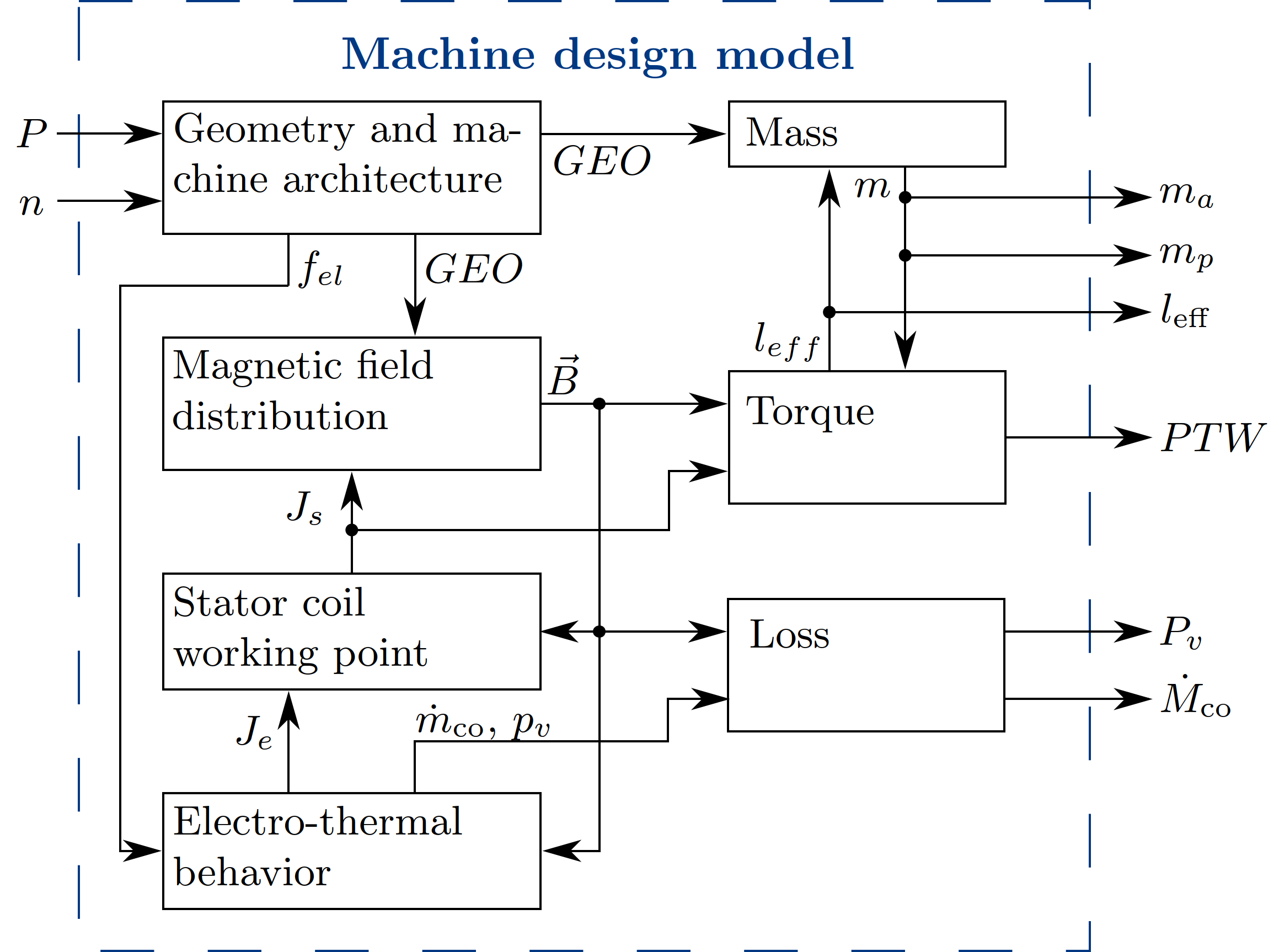}
	\caption{Calculation procedure of the analytical design model for synchronous radial flux machines with the requirements as an input on the left side and target parameter as an output on the right side.}    
	\label{fig:calc_procedure_RFM} 
\end{figure}

Superconducting synchronous radial flux machines can be generally categorized into partially and fully superconducting machines (SCM). Partially superconducting machines are in turn subdivided into machines with a superconducting DC rotor and a normal conducting AC stator or into machines with a normal conducting AC rotor and a superconducting DC stator. The former topology of the partial SCM is investigated in this paper. In contrast, the topology of the fully SCM uses superconducting DC coils in the rotor and superconducting AC coils in the stator. In superconducting DC rotor coils, only low losses occur, which have a minimal effect on the current-carrying capacity \cite{Barnes.2005}. The current-carrying capacity of DC coils is limited in general by their self field and mechanical limitations. 
However, in AC stator coils strong and fast varying electromagnetic fields are present and a substantial amount of losses occur. Due to the strongly temperature-dependent electric characteristics of superconductors, they have to be coupled to thermics and flow dynamics of the cooling process.\\
To optimize a large number of parameters of a machine in a reasonable time, an analytical model is developed.
Figure~\ref{fig:rfm_model_cold_airgap} shows the general scheme.
The main features of the machine model are as follows. Firstly, the model supports the design of synchronous radial flux machines with air gap windings. Secondly, the rotor is free of iron and the stator is enclosed by a yoke of iron. Regardless of the machine topology, the stator iron is operated at ambient temperatures. This has several reasons as investigated in \cite{Liu.2018}. Firstly, this is advantageous for the relative permeability which increases at cryogenic temperatures, but also disadvantageous in the losses which also increase. The stator iron core is operated close to the saturation point of the yoke material. This leads to high iron losses which would burden the cryogenic system heavily. The iron loss is determined according to \cite{Bertotti.1985} and \cite{Bertotti.1985b}. Finally, the variability in geometry, rotor excitation, and materials allow the optimization of the machine KPIs with respect to a large number of model parameters. The varied internal parameters are summarized in Table~\ref{tab:optimisation_parameters}.\\
\begin{table}[!h]
	\caption{\label{tab:optimisation_parameters}Main parameters of the machine design model}
	\begin{center}
		\begin{tabular}{lll}
			\hline
			Symbol 			& Unit 				& Explanation \\
			\hline
			$p$				& $\mathrm{-}$	& Number of pole pairs \\
			$f_{el}$		& $\mathrm{Hz}$	& Electrical frequency \\
			$m$				& $\mathrm{-}$	& Number of phases \\
			$I_{n,\mathrm{max}}$		& $\mathrm{-}$	& Maximum normalized current in the stator \\
			$r_{ri}$		& $\mathrm{m}$	& Inner radius of the rotor \\
			$\alpha_{0}$	& $\mathrm{\%}$	& Duty cycle of the stator coils \\
			$d_{s,c}$		& $\mathrm{m}$	& Thickness of the stator coils \\
			$B_{y,\mathrm{sat}}$		& $\mathrm{T}$	& Saturation flux density of the yoke \\
			$f_{\mathrm{csr}}$		& $\mathrm{-}$	& Coil support ratio \\
			$J_{r}$			& $\mathrm{A \, mm^{-2}}$		& Current density in rotor coils \\
			$d_{m}$			& $\mathrm{m}$	& Thickness of rotor coils \\
			$\alpha_{2}$	&  $\mathrm{^\circ}$	& Inner angle of rotor coils \\
			$\alpha_{3}$	&  $\mathrm{^\circ}$	& Angle between adjacent rotor coils \\
			$\theta_{\mathrm{load}}$ &  $\mathrm{^\circ}$	& Load angle \\
			\hline
		\end{tabular}
	\end{center}
\end{table}
The calculation procedure is presented in Figure~\ref{fig:calc_procedure_RFM}. The main part of the machine design model is the determination of the machine geometry $GEO$ and its architecture. External requirements are the power $P$ and the rotation speed $n$. The electro-thermal behavior of a specific superconductor is calculated in dependence of the electrical frequency $f_{el}$, the flux density in stator coil area $B_s$, and the normalized current $I_n$ at the calculated working temperature $T$. This approach allows to design a machine with a specific engineering current density $J_e$ and computes the required cryogenic cooling conditions. It is presented in more detail in Section~\ref{electro_thermal_model}. The magnetic field that penetrates the superconductors in the stator winding is mainly generated by the rotor and by the adjacent stator coils.\\
The geometry forms the basis to calculate the magnetic field inside the machine, presented in Section~\ref{electromagnetic}. An exact calculation of the field distribution is essential to the machine design process caused by the nonlinear electro-thermal behavior of the superconductor. Furthermore, the field information is indispensable in the calculation of the mean torque $\overline{T}$ and the cryogenic coolant consumption $\dot{M}_{co}$. These are computed by the local mass flow densities $\dot{m}_{co}$ and loss densities $p_v$ that are generated by the electro-thermal behavior. The target torque of the machine is reached by adjusting the effective machine length $l_{\mathrm{eff}}$ for a given geometry. Active parts, cryogenic and mechanic support structures as well as housing are part of the geometry and are described in detail in Section~\ref{mechanic_cryogenic}. These parts can consist of different materials depending on their respective functionality. The active mass $m_a$ and passive mass $m_p$ of the machine are calculated with this approach. Moreover, the entire geometry is adapted to the machine architecture.\\
In the model, the following effects are neglected. Firstly, due to the two-dimensional static magnetic field calculation, end effects are not considered. Secondly, for the calculation of the field, the thickness and the relative permeability of the stator yoke are assumed to be infinite. Thereby its hysteresis behavior is not considered. To estimate the thickness of the stator yoke, the radial field component of the total field at the transition to the yoke is integrated over a pole and then divided by the saturation flux density $B_{y,sat}$. Finally, the current density distribution in the coils is assumed to be constant. Therefore, the skin effects in normal conductors and shielding in superconductors are neglected. This approach is reasonable because a coil consists of several windings with a constant current.\\
The working point current density $J_{wp}$ of the stator coil is defined at equilibrium between the spatially highest AC losses and the cooling conditions. Therefore, the current density in the stator is varied until the AC losses in presence of the total electromagnetic field meet this condition.

\subsection{Electromagnetic design}
\label{electromagnetic}
The machine design requires the exact magnetic field information in the entire stator region. This is necessary to determine the local loss density accurately and with spatial resolution. Due to the cylindrical symmetry of the machine model, the calculation of the electromagnetic fields can be simplified and is done in the frequency domain \cite{Woodson.1966}. The fields are calculated for each coil system with identical phase angle or current density, and are subsequently superimposed.\\
A vector potential $\Phi$ is used to calculate the field generated by the coils. This potential has only a $z$-direction, if the field distribution is solved in the $x$-$y$-plane and the current density $J$ has only a $z$-component. The Laplace equation~(\ref{eq:laplace_coils}) allows the calculation of the magnetic field $H$ outside of the coil area in the regions $m$, shown in Figure~\ref{fig:rfm_model_cold_airgap}. 

\begin{equation}
\frac{\partial^2 \Phi_{z,m}}{\partial r^2} + \frac{1}{r} \frac{\partial \Phi_{z,m}}{\partial r} + 	\frac{1}{r^2} \frac{\partial^2 \Phi_{z,m}}{\partial \theta^2} = 0
\label{eq:laplace_coils}
\end{equation}

The Poisson equation~(\ref{eq:poisson_coils}) is used to solve the field inside the coils with the current density $J_{z,m}$ and the vacuum permeability $\mu_0$.

\begin{equation}
\frac{\partial^2 \Phi_{z,4}}{\partial r^2} + \frac{1}{r} \frac{\partial \Phi_{z,4}}{\partial r} +\frac{1}{r^2} \frac{\partial^2 \Phi_{z,4}}{\partial \theta^2} = -\mu_{0} \, J_{z,m}
\label{eq:poisson_coils}
\end{equation}

The Laplace equation and Poisson equation can be solved for the $n^{th}$ harmonic by the general solutions given in Equation~(\ref{eq:gs_coils_laplace_source}) by Liu \cite{Liu.2018b}, including the singularity at $np=2$. The Fourier coefficients are $A_{n,m}$, $C_{n,m}$ and $J_{n,m}$, if a current flows in the specific area.

\begin{equation}
\Phi_{z,m}\left(r,\theta\right) =
\begin{cases}
\!\begin{aligned}
& \sum_{n \, \mathrm{odd}} \left(  A_{n,m} \, r^{np} + C_{n,m} \, r^{-np} \right. \\[5pt]
& \left. - \frac{J_{n,m} \, r^2}{4-(np)^2}\right) \cos(np\theta) \\[5pt]
\end{aligned} & \text{if } np \neq 2\\[5pt]
\!\begin{aligned}
& \sum_{n \, \mathrm{odd}} \left( A_{1,4} \, r^2 + C_{1,4} \, r^{-2} \right. \\[5pt]
& \left. - \frac{1}{4} \, J_{1,m} r^2 	\ln(r)\right) \cos(2\theta) \\[5pt]
\end{aligned} & \text{if } np = 2	
\end{cases}
\label{eq:gs_coils_laplace_source}
\end{equation}

The magnetic flux density $\vec{B}$ is calculated as the curl of the vector potential. The Fourier component $J_{n,m}$ of the $n^{th}$ harmonic in the $m^{th}$ region is calculated as: 

\begin{equation}
J_{n,m} =
\begin{cases}
\!\begin{aligned}
&  J_{max} \left( \, \frac{-2 \sin(\alpha_3 n) + 2 \sin(n(\alpha_3 - \pi))}{n \, pi} \right. \\[5pt]

& \left. + \frac{4 \cos(\alpha_2 n) \sin\left(\frac{n \pi}{2}\right)}{n \, pi} \right)\\[5pt]
\end{aligned} & \text{if rotor}\\[5pt]
\!\begin{aligned}
2 \, \alpha_0 \, J_{max} \, \sinc\left( \frac{n \, \alpha_0}{2} \right) \\[5pt]
\end{aligned} & \text{if stator}	
\end{cases}
\label{eq:fourier}
\end{equation}

By setting boundary conditions for the tangential magnetic field and radial flux density at the coordinate origin, between adjacent regions and in the inner yoke radius, the Fourier coefficients $A_{n,m}$ and $C_{n,m}$ can be calculated.

\subsection{Mechanical and cryogenic design}
\label{mechanic_cryogenic}
The thickness of the magnetic air gap $d_{mag}$ influences the field distribution in the machine and thus also the torque and the losses in the stator windings.
\begin{figure}[!h!]
	\begin{center} 
		\subfloat[Fully superconducting machine\label{subfig:airgap_fullySC}]{
			\def\svgwidth{0.47\textwidth}    
			\includegraphics{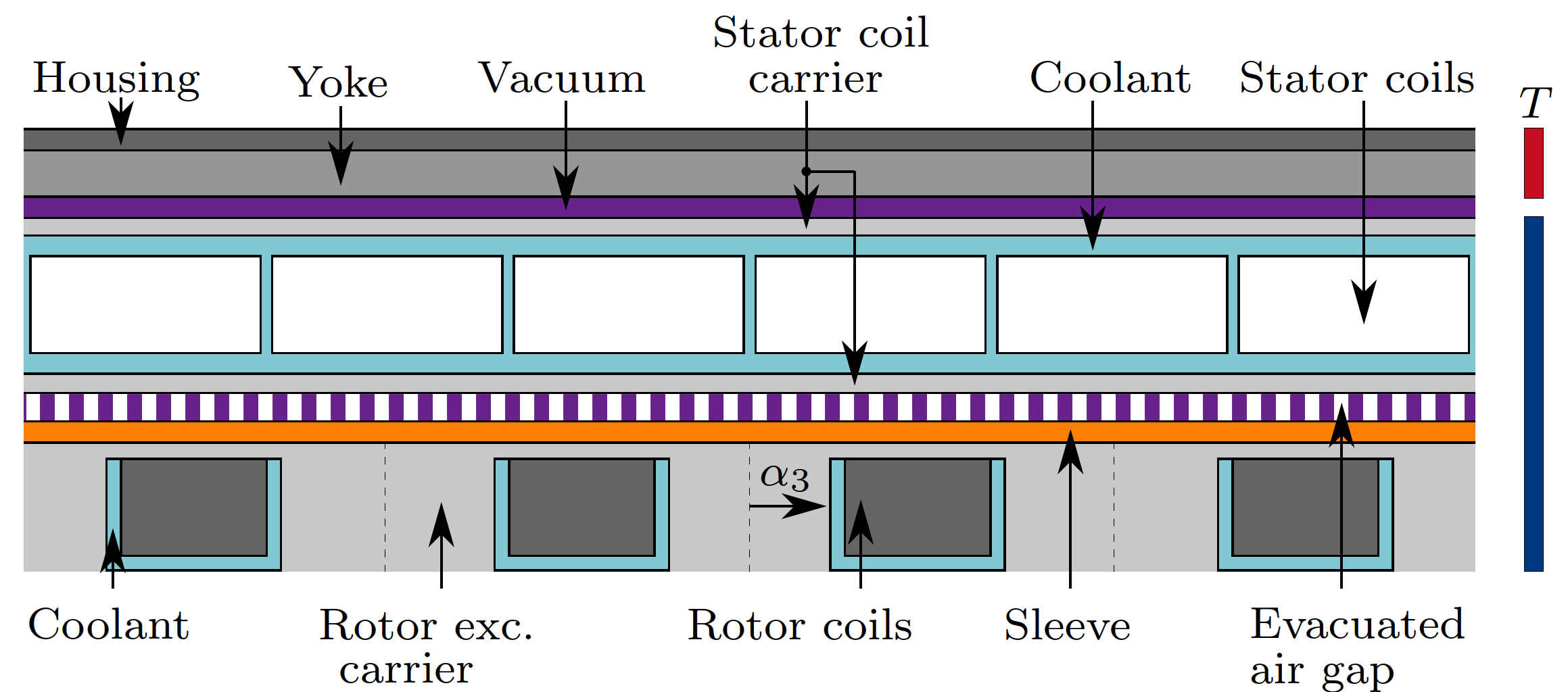}
		}
		\vskip 0.005\textwidth
		\subfloat[Partially superconducting rotor machine\label{subfig:airgap_partiallySC_rotor}]{       
			\def\svgwidth{0.47\textwidth}    
			\includegraphics{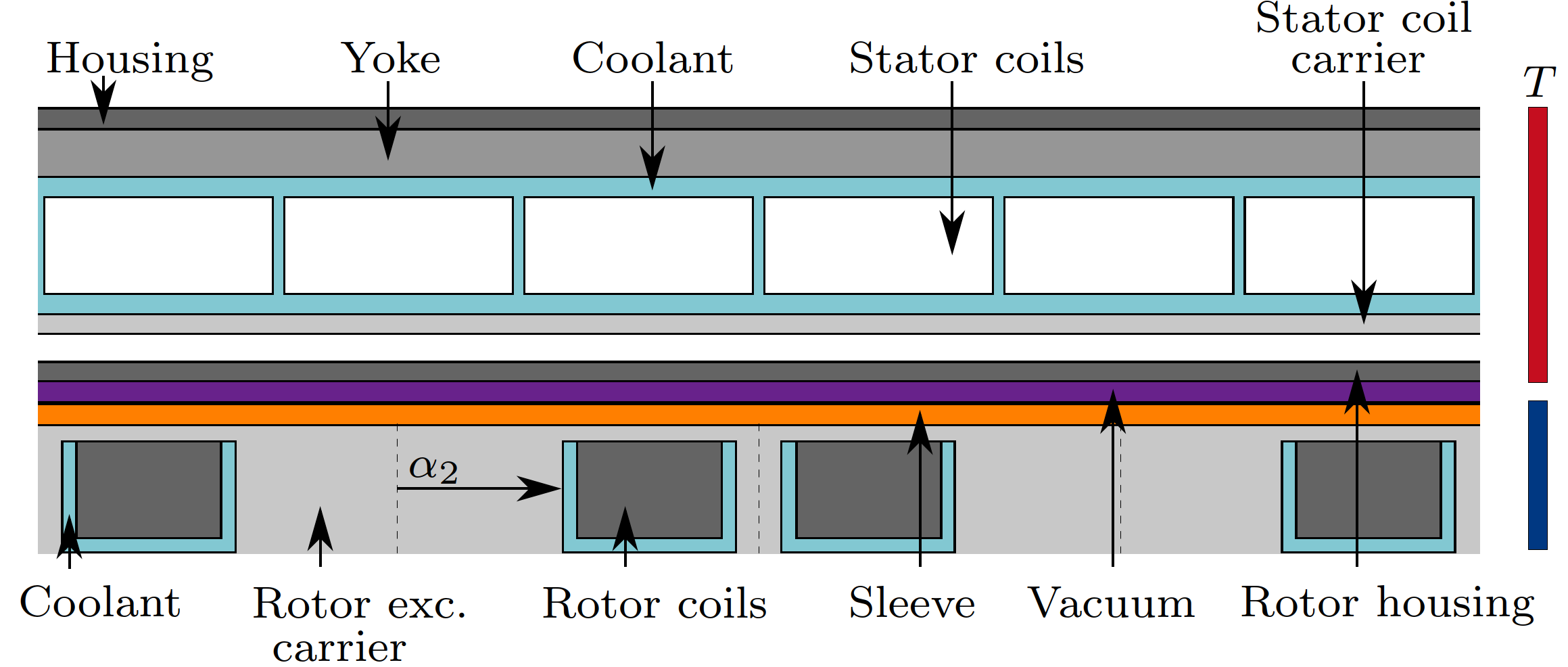}
		}
		\caption{Detailed air gap design of a fully superconducting machine \protect\subref{subfig:airgap_fullySC}, partially superconducting rotor machine \protect\subref{subfig:airgap_partiallySC_rotor} and temperature areas are marked for cryogenic \legendsquare{fill=blue} and ambient temperature \legendsquare{fill=red}.
		}
		\label{fig:airgap}
	\end{center}
\end{figure}
Therefore, its radial thickness has to be determined accurately within the thermal and mechanical boundaries of the machine.\\
The magnetic air gap consists of the sleeve and several cryo walls with the corresponding thicknesses $d_{sl}$ and $d_{cw}$, depending on the machine topology, shown in Figure~\ref{fig:airgap}. To take into account manufacturing tolerances and displacements caused by rotor dynamics, an additional air gap thickness of $d_{ag} = 1 \, \mathrm{mm}$ will be considered. Furthermore, a thickness of $d_{va} = 1 \, \mathrm{mm}$ is assumed for the vacuum to prevent heat conduction into the cryo system. The magnetic air gap excluding the stator coils is calculated as follows:
\begin{equation}
r_{si}-r_{ro} = d_{mag} = \sum d_{cw} + d_{sl} + d_{ag} + d_{va}	
\end{equation}
The cryo walls labeled as rotor housing and stator housing provide thermal insulation. These can either be loaded under external or internal pressure and thus buckling should be excluded by a sufficient thickness of the cryo walls $d_{cw}$. The calculation of the thickness is done according to \cite{VerbandderTUVe.V.2000} and \cite{VerbandderTUVe.V.2006}. A special variant occurs in a partially superconducting machine. In the rotor, an additional centrifugal force occurs in the cryo wall, which is taken into account in its design.\\
An important mechanical part is the sleeve which supports the rotor carrier and the rotor coils due to the centrifugal forces. Its thickness is determined through an analytical press-fit model which is used to assess the static strength of the sleeve. The rotor is modeled as a compound of three adjacent hollow cylinders which represent the rotor carrier, the rotor coils, and the sleeve. Thus, the homogenized tangential stress in the sleeve is calculated as a result of its radial overclosure. The displacements due to the inner pressure, the centrifugal forces, and the thermal expansion are considered. The calculation for the stress components and the radial displacement of rotating hollow cylinders is based on \cite{Boresi.2003} and \cite{Eslami.2013}. Finally, the thickness of the sleeve  $d_{sl}$ is calculated iteratively by variation of itself and the radial overclosure. It strongly depends on the rotation speed, the rotor radius as well as the coil thickness and was internally cross-checked by FEM.\\
The thickness of the machine housing is designed such that the torque of the machine can be transmitted. By designing these mechanical and thermodynamical support structures, the passive mass of the machine is calculated. The active mass includes the yoke and the stator and rotor coils including the winding head. Both stator coils and rotor coils are assumed to be racetrack coils.

\begin{figure}[!h]
	\centering       
	\def\svgwidth{0.33\textwidth}    
	\includegraphics{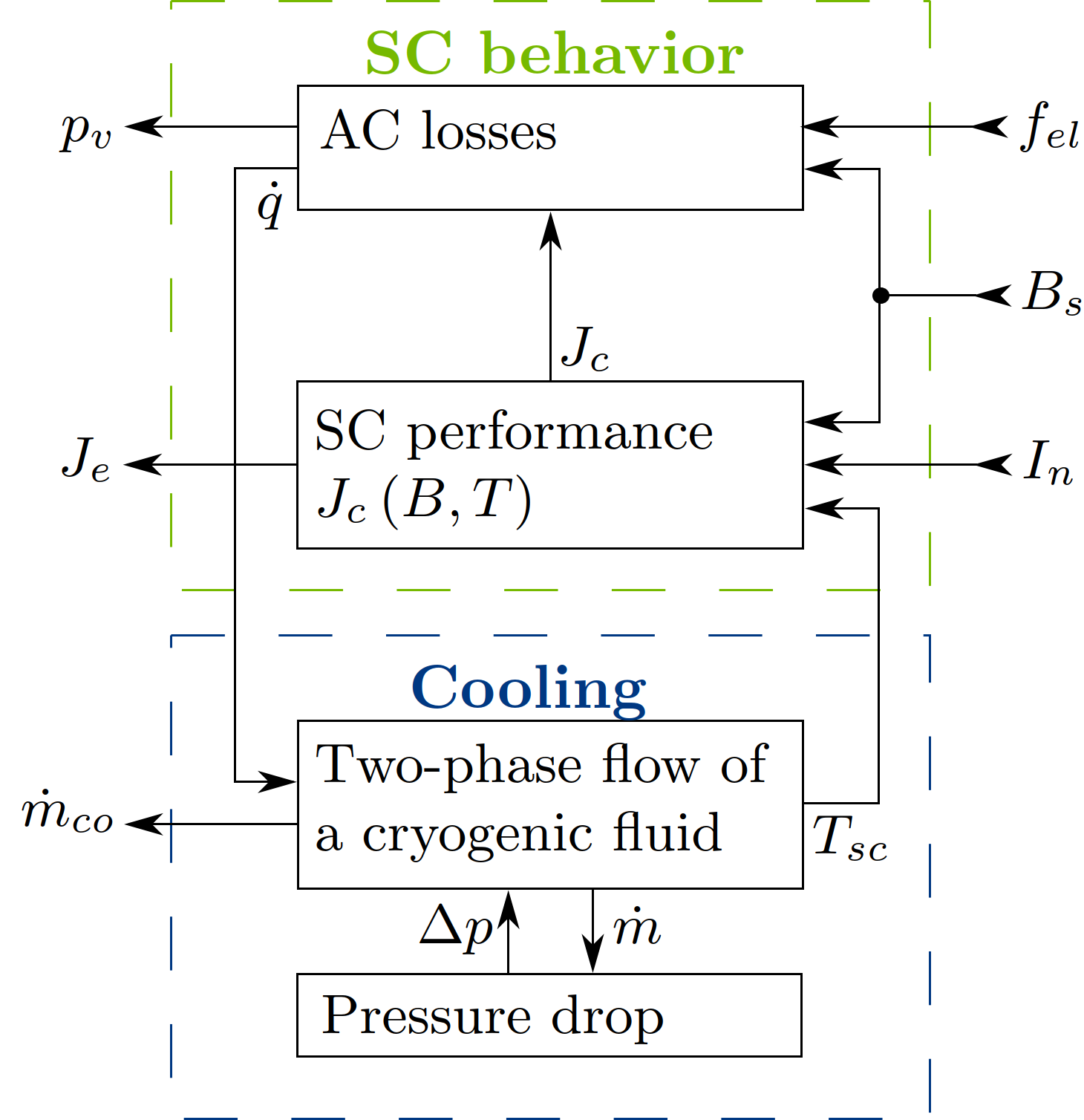}
	\caption{The calculation procedure of the electro-thermal model is divided into the behavior of the superconductor (SC behavior) and the cooling procedure (Cooling).}    
	\label{fig:calc_procedure_ETM} 
\end{figure}

\subsection{Electro-thermal model}
\label{electro_thermal_model}

At the beginning of the machine design, the current density in the stator coils is not known. Due to the amount of AC loss compared to the cooling capability of the liquid cooling, the electric and thermal behavior of the stator windings has to be treated in a coupled closed-loop model which is schematically depicted and presented in Figure~\ref{fig:calc_procedure_ETM}. Initially, the computed engineering current density is independent of the machine design and it is calculated as a function of an alternating external field penetrating a wire that conducts a transport current. Such an electro-thermal model is developed for superconducting MgB\textsubscript{2} wires and normally conducting copper litz wires.\\
For the cooling of MgB\textsubscript{2} wires, a two-phase flow is assumed. This cooling mechanism is modeled by empirical equations according to \cite{GesellschaftVerfahrenstechnikundChemieingenieurwesen.2013}. Furthermore, the two-phase cooling concept is required to keep the MgB\textsubscript{2} temperature as low as possible in the case of high AC loss. The calculation procedure is shown in Figure~\ref{fig:calc_procedure_ETM} and is divided into two parts, the behavior of the superconductor (SC behavior) and cooling. Besides the wire geometry and the normalized current $I_n$ of the superconductor, the stator coil flux density $B_{s}$ and the electrical frequency $f_{el}$ are the input parameters of the model. The SC performance was determined by measurements of the critical current in a wide range of magnetic fields and temperatures. AC losses are divided into magnetization loss, eddy current loss, and coupling current loss. They are calculated according to \cite{Bean.1962}, \cite{Wilson.1986} and \cite{Oomen.2000}. The magnetization and eddy current losses were cross-checked with Comsol and the results show that the loss we obtain analytically is slightly higher compared to FEM. The loss creates locally specific heat flux densities $\dot{q}$ depending on the material and temperature in the different parts of the wire.
\begin{table}[!h]
	\caption{\label{tab:optimisation_parameters_example}Machine parameters for the hybrid-wing-body concept aircraft N3-X \cite{Felder.2011}}
	\begin{center}
		\begin{tabular}{llll}
			\hline
			Parameter		& Unit 					& Fully SCM		&  Partially SCM\\
			\hline
			Stator winding& $\mathrm{-}$			& MgB\textsubscript{2} & Cu \\
			Rotor winding	& $\mathrm{-}$			& HTS			& HTS \\
			Cooling liquid	& $\mathrm{-}$			& LH\textsubscript{2} & Novec 7500 \\
			$p$				& $\mathrm{-}$			& $3$ - $10$	& $3$ - $10$ \\
			$f_{el}$		& $\mathrm{Hz}$			& $225$ - $750$		& $225$ - $750$	\\
			$m$				& $\mathrm{-}$			& $3$			& $3$ \\
			$I_{n,\mathrm{max}}$		& $\mathrm{-}$			& $0.7$			& - \\
			$r_{ri}$		& $\mathrm{m}$			& $0.14$ - $0.18$	& $0.12$ - $0.2$ \\
			$\alpha_{0}$	& $\mathrm{\%}$			& $95$			& $95$ \\
			$d_{s,c}$		& $\mathrm{mm}$			& $6$ - $11$	& $14$ - $24$ \\
			$B_{y,\mathrm{sat}}$		& $\mathrm{T}$			& $2.3$			& $2.3$ \\
			$f_{\mathrm{csr}}$		& $\mathrm{-}$			& $1/3$			& $1/3$ \\
			$J_{r}$	& $\mathrm{A \, mm^{-2}}$		& $300$ 		& $300$\\
			$d_{m}$			& $\mathrm{mm}$			& $12$			& $12$, $24$\\
			$\alpha_{2}$	& $\mathrm{^\circ}$		& $60$ - $70$	& $40$ - $65$ \\
			$\alpha_{3}$	& $\mathrm{^\circ}$		& $0.04$ - $0.051$	& $0.034$ - $0.057$ \\
			$\theta_{\mathrm{load}}$ & $\mathrm{^\circ}$		& $90$			& $90$ \\
			Calculated SCM	& $\mathrm{-}$			& $1440$		& $2880$ \\
			\hline
		\end{tabular}
	\end{center}
\end{table}
The basic assumption of two-phase cooling is that this emitted heat over the insulated cable surface must be equal to the heat output during the evaporation of LH\textsubscript{2}. As a result, taking into account the pressure drop, the maximum temperature $T_{sc}$ inside the conductor can be calculated whereby the critical current is adjusted. This calculation is performed in a loop until the difference between two iteration steps falls below $0.1 \, \mathrm{K}$ to determine the static temperature. Results are the engineering current density $J_e$, which is related to the sum of conductor area and cooling area, the mass flow density $\dot{m}_{co}$ and the loss density $p_v$.\\
A similar procedure exists in the electro-thermal calculation of a copper litz wire with a single-phase oil cooling. The aim of this calculation is to determine the maximum temperature $T_{max}$ at a given engineering current density $J_e$ inside the litz wire considering an inlet temperature $T_{in}$ of the cooling liquid. It is assumed that the litz wire is cast in resin and cooled by Novec 7500 \cite{3MDeutschlandGmbH.2014} in a channel from one side. The AC loss in the copper filaments are calculated according to \cite{Lammeraner.1966}, which are divided into skin effects, proximity effects and additionally the ohmic loss. Subsequently, the fluid dynamics can be considered according to \cite{GesellschaftVerfahrenstechnikundChemieingenieurwesen.2013}.\\
Since the electromagnetic field in the stator winding was computed spatially dependent the loss density among the coil vary locally as well. Therefore, the electro-thermal model is evaluated in a mesh with a radial resolution of $1 \, \mathrm{mm}$ and a tangential resolution of $200$ steps per coil. The total AC loss $P_{v}$ is computed by the integration of these local loss densities. Furthermore, local temperature hot spots are visible and can be considered in the design of the cooling. The efficiency $\eta$ of the machine takes into account iron losses and stator losses.

\section{Analysis for N3-X motor requirements}

\subsection{Description}

We used our model to analyze the potential of a fully SCM and partially SCM for the following high-level requirements: P = $\mathrm{3 \, MW}$ and n = $\mathrm{4500 \, rpm}$. These had been derived from aircraft design considerations for the hybrid-wing-body concept aircraft N3-X \cite{Kim.2016} which is powered by a turboelectric distributed propulsion system incorporating 15 motors with the given power and speed.\\
We assumed the following materials for the machine design. In the stator winding of the fully SCM we assume a 114-multifilament MgB\textsubscript{2} wire \cite{Wan.2017} which is cooled with liquid hydrogen. The $J_c \left(B,T\right)$ characterization of this wire was carried out experimentally in a field range of $0$ to $2 \, \mathrm{T}$ and a temperature range of $20$ to $33 \, \mathrm{K}$. In the stator winding of the partially SCM we assume a copper litz wire with a filament diameter of $0.5 \, \mathrm{mm}$ which is cooled by the silicon oil Novec 7500. The maximum allowable temperature of the insulated wire is $180 \, \mathrm{^\circ C}$. For both topologies the rotor field is generated either by a HTS single pancake coils $\left( d_m = 12 \, \mathrm{mm} \right)$ or a HTS double pancake coil $\left( d_m = 24 \, \mathrm{mm}\right)$ with a rotor current density $J_r$ of $300 \, \mathrm{A \, mm^{-2}}$ at $25 \, \mathrm{K}$ \cite{Oomen.03.09.2019}. As the yoke sheet metal, the commercially available soft magnetic cobalt-iron alloy Vacodur \cite{Vacuumschmelze.2016} is assumed. A titanium alloy \cite{SpecialMetalsCorporation.2007} and an aluminum alloy \cite{ASMHandbook.1990} are assumed as the material of the sleeve and cryo walls, respectively. Both alloys combine high strength and an insensitive hydrogen permeability \cite{Robertson.1977} \cite{Walter.1973}. The material of the housing is assumed to be titanium \cite{Donachie.2000}.\\
Besides the number of pole pairs we varied internal geometry parameters of the machines according to Table~\ref{tab:optimisation_parameters_example}, to find the configurations with the highest power-to-weight ratios and efficiencies. As the models run very fast, we did not use a dedicated optimization algorithm but calculated all configurations that can be generated by permutating the parameters. 

\subsection{Results}

\begin{figure}
	\subfloat[All results of the fully SCM
	\label{subfig:PTW_Bwp_all_MgB2}]{
		\centering
		\def\svgwidth{0.33\textwidth}    
		\includegraphics{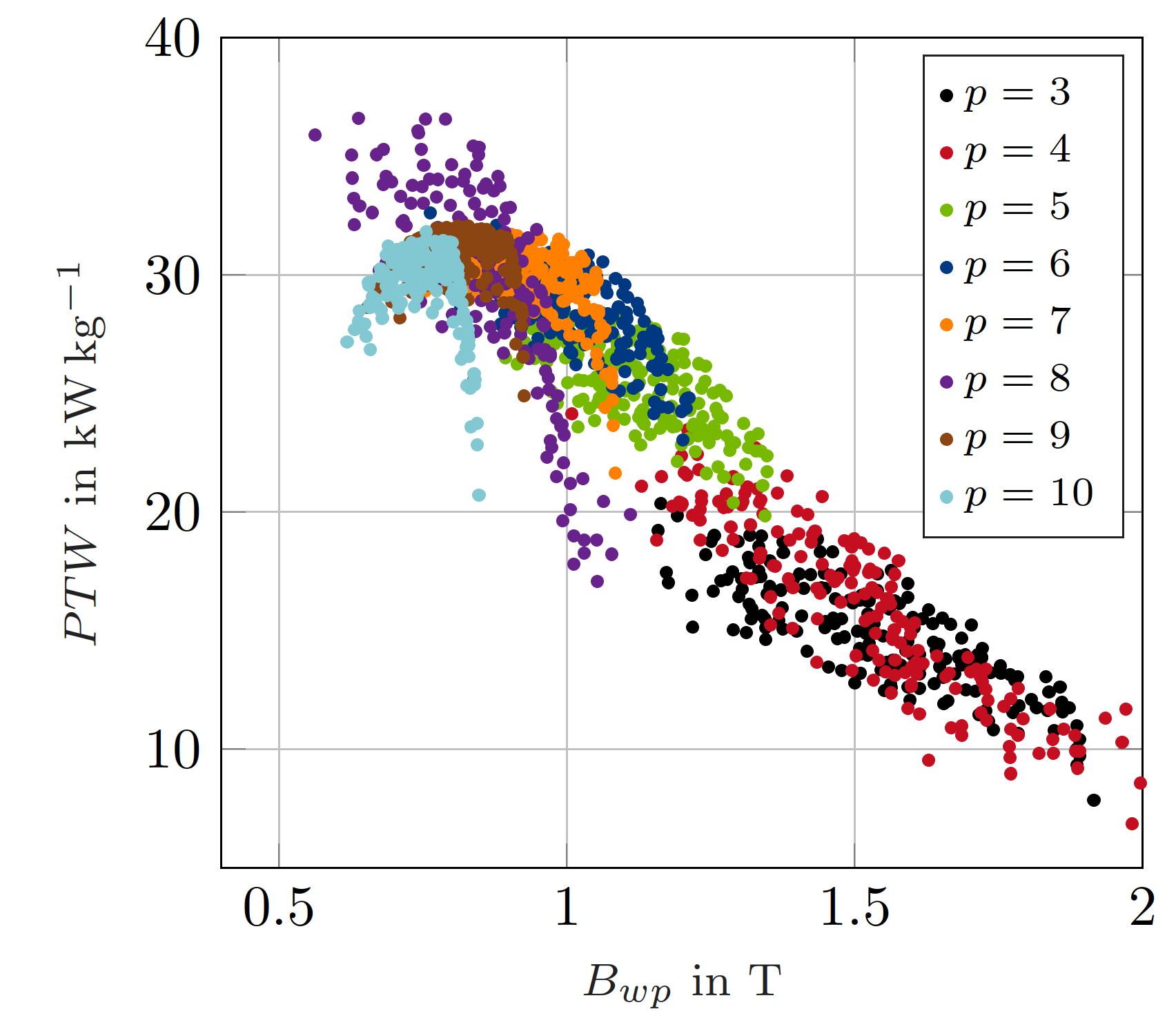}
	}
	\subfloat[All results of the partially SCM
	\label{subfig:PTW_Bwp_all_Cu}]{
		\centering
		\def\svgwidth{0.33\textwidth}    
		\includegraphics{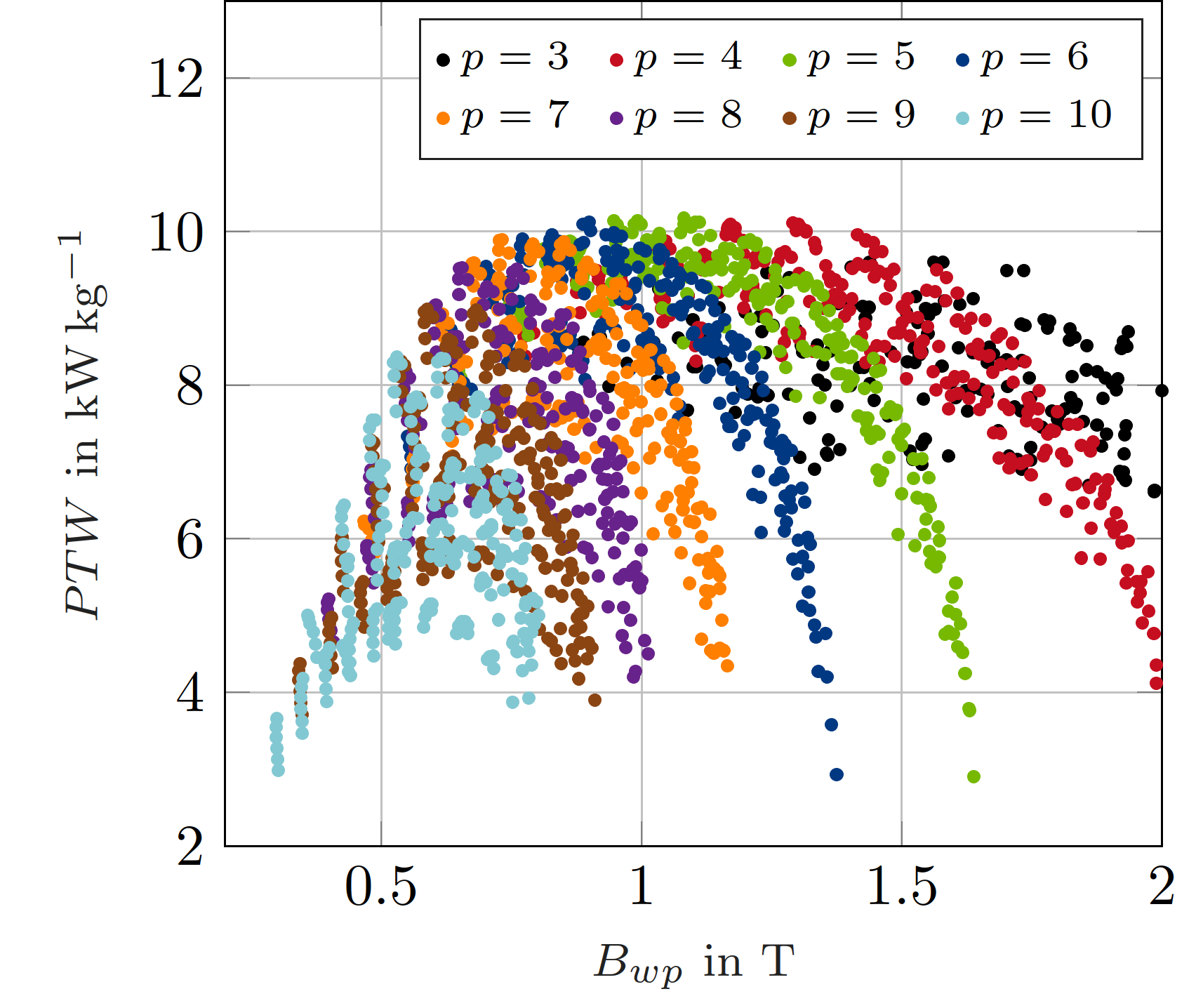}
	}
	\vskip 0.00\textwidth
	\subfloat[All results of the fully SCM
	\label{subfig:PTW_Jwp_all_MgB2}]{
		\centering
		\def\svgwidth{0.33\textwidth}    
		\includegraphics{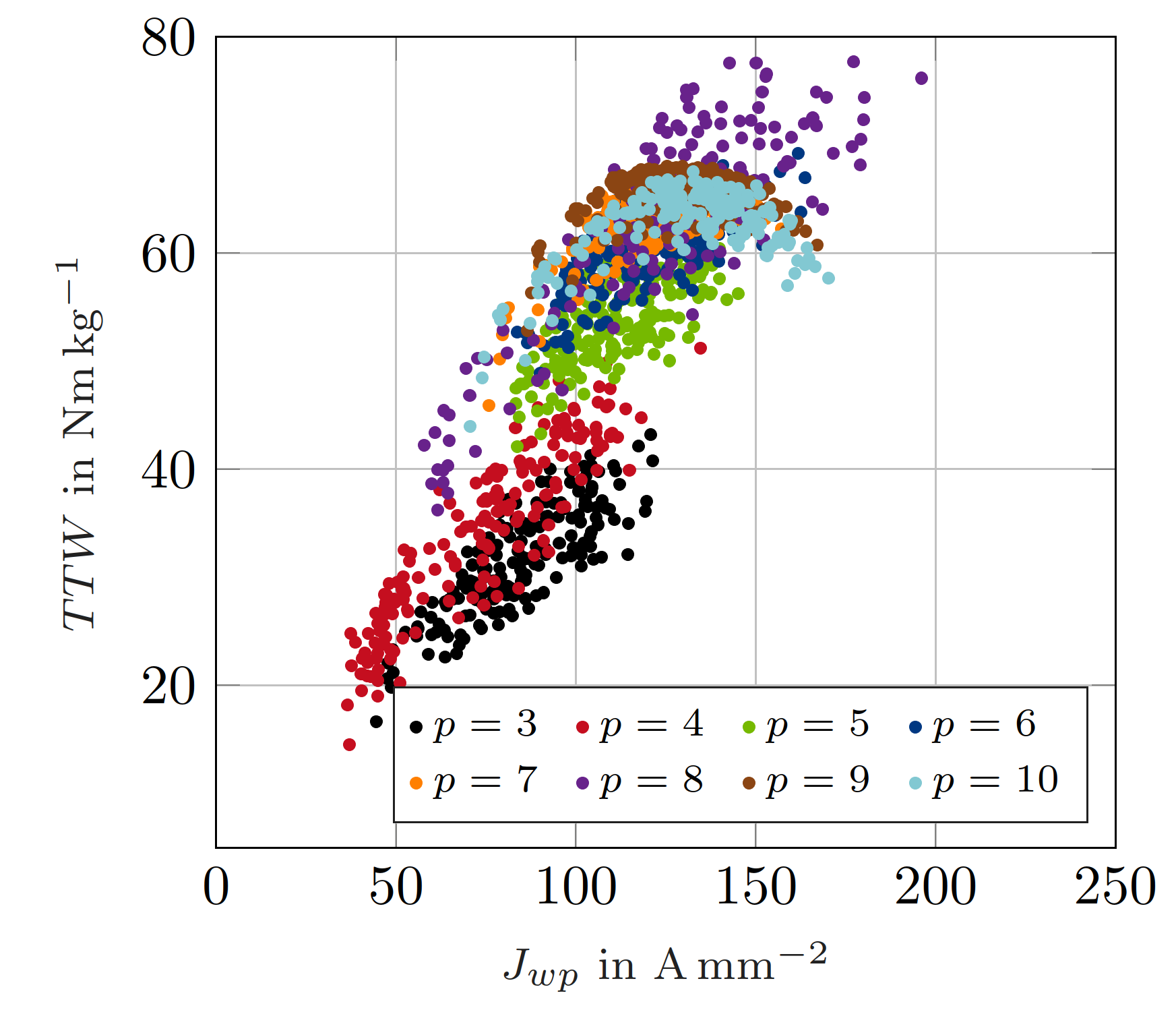}
	}
	\subfloat[All results of the partially SCM
	\label{subfig:PTW_Jwp_all_Cu}]{
		\centering
		\def\svgwidth{0.33\textwidth}    
		\includegraphics{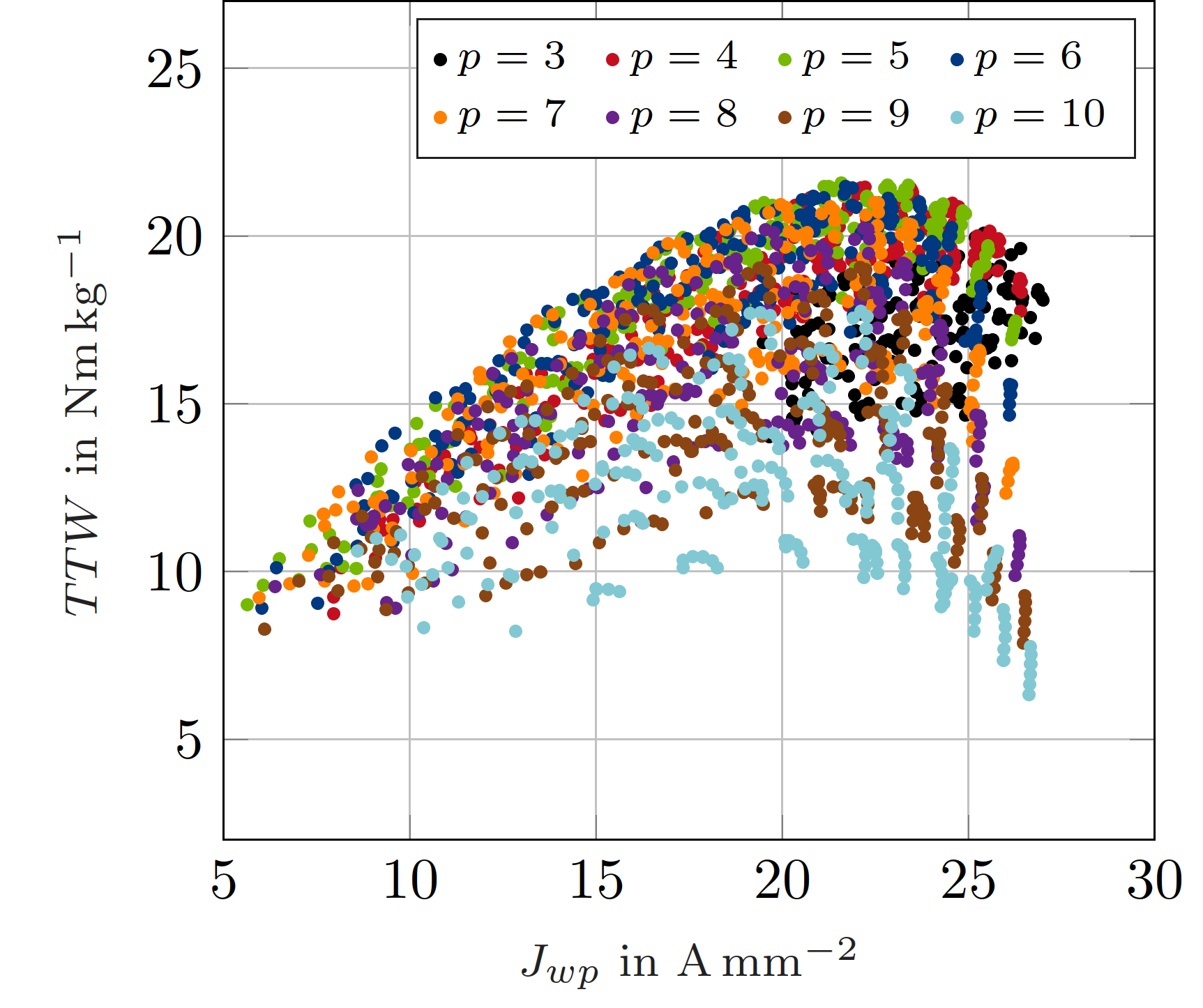}
	}
	\vskip 0.00\textwidth
	\caption{Power-to-weight ratio $PTW$ as a function of the working point flux density $B_{wp}$ in \protect\subref{subfig:PTW_Bwp_all_MgB2}, \protect\subref{subfig:PTW_Bwp_all_Cu} and torque-to-weight ratio $TTW$ as a function of the working point current density $J_{wp}$ in \protect\subref{subfig:PTW_Jwp_all_MgB2}, \protect\subref{subfig:PTW_Jwp_all_Cu} for pole pair numbers $p$ between $3$ $(225\, \mathrm{Hz})$ and $10$ $(750\, \mathrm{Hz})$.}
	\label{fig:results_PTW_coil}
\end{figure}

Varying the parameters as presented in Table~\ref{tab:optimisation_parameters_example} results in 4320 computed configurations which are shown in Figure~\ref{fig:results_PTW_coil}. The variation parameters include the number of pole pairs $p$, the inner radius of the rotor $r_{ri}$, the thickness of the stator coils $d_{s,c}$, the thickness of the rotor coils $d_m$ and the inner angle of the rotor coils $\alpha_{2}$.\\
In Figure~\ref{subfig:PTW_Bwp_all_MgB2} and \ref{subfig:PTW_Bwp_all_Cu}, the power-to-weight ratio $PTW$ is shown as a function of the working point flux density $B_{wp}$ for different number of pole pairs for the fully SCM and partially SCM, respectively. The power-to-weight ratio $PTW$ includes the active and passive mass of the machines. For both topologies, it can be seen that the $PTW$ increases with increasing $p$ up to a certain number of pole pairs. While for fully SCM the lightest designs with $PTW > 30 \, \mathrm{kW \, kg^{-1}}$ can be found for pole pair number between 6 and 10, the lightest partially SC machines with $PTW > 10 \, \mathrm{kW \, kg^{-1}}$ can be found for pole pair numbers between 4 and 6. Taking a closer look at fully SC machine designs with pole pair number 8 reveals that its $PTW$ values are outstanding (compared to p = 7 and p = 9) with the highest value of $36.6 \, \mathrm{kW \, kg^{-1}}$.\\
The optimum design range concerning the working point flux density and the pole pair number is considerably smaller for fully SCM than for partially SCM being even more pronounced at high $PTW$. This behavior can be attributed to the high sensitivity of the MgB\textsubscript{2} wire to AC loss. The highest $PTW$ values are reached for working point flux densities in the range of $0.55 \, \mathrm{T}$~-~$0.9 \, \mathrm{T}$ for fully SCM and in the range of $0.8 \, \mathrm{T}$ - $1.4 \, \mathrm{T}$ for partially SCM. In the case of a small number of pole pairs, the $PTW$ increases mainly due to the reduction of the yoke thickness and higher electric frequencies that are directly proportional to the pole pair number. Both effects saturate with an increasing number of pole pairs. The higher electric frequencies enhance the generation of AC loss and consequently, the engineering current density has to be reduced leading to more active material in the stator. Thus, an optimum of the $PTW$ can be found. In the case of the partially SC machine, the frequency-independent ohmic losses dominate the total losses in the copper litz wires. This leads to a lower sensitivity of $PTW$ of the partially SCM to the number of pole pairs and the magnetic field $B_{wp}$ at the working point of the stator.

\begin{figure}
	\begin{center}
		\hspace{-1.75cm}
		\subfloat[Stator current density\label{subfig:Jphases_theta}]{
			\centering
			\def\svgwidth{0.33\textwidth}
			\includegraphics{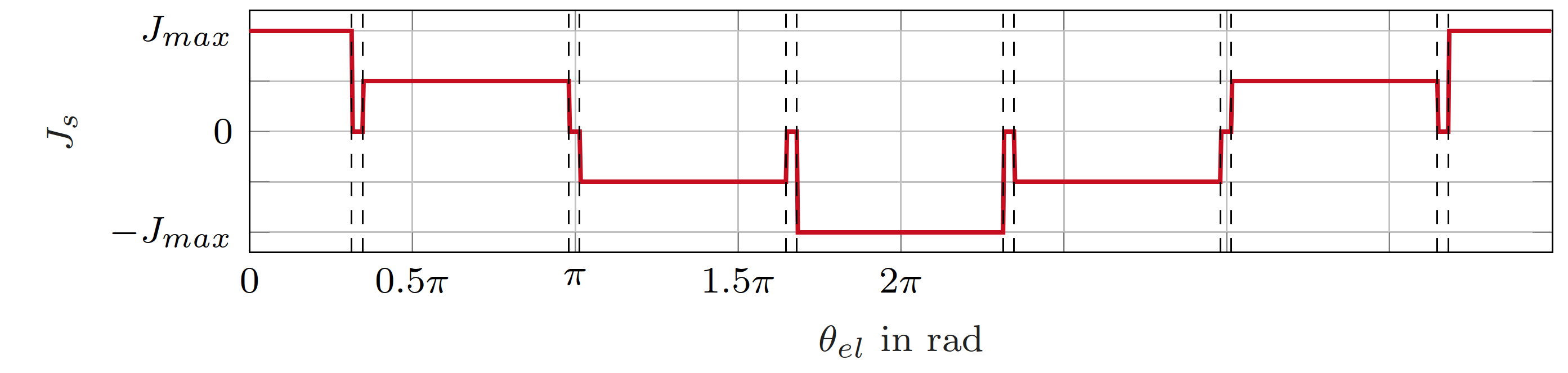}
		}
		\vskip 0.0\textwidth
		\subfloat[Flux density\label{subfig:B_r_theta}]{
			\centering
			\def\svgwidth{0.33\textwidth}
			\includegraphics{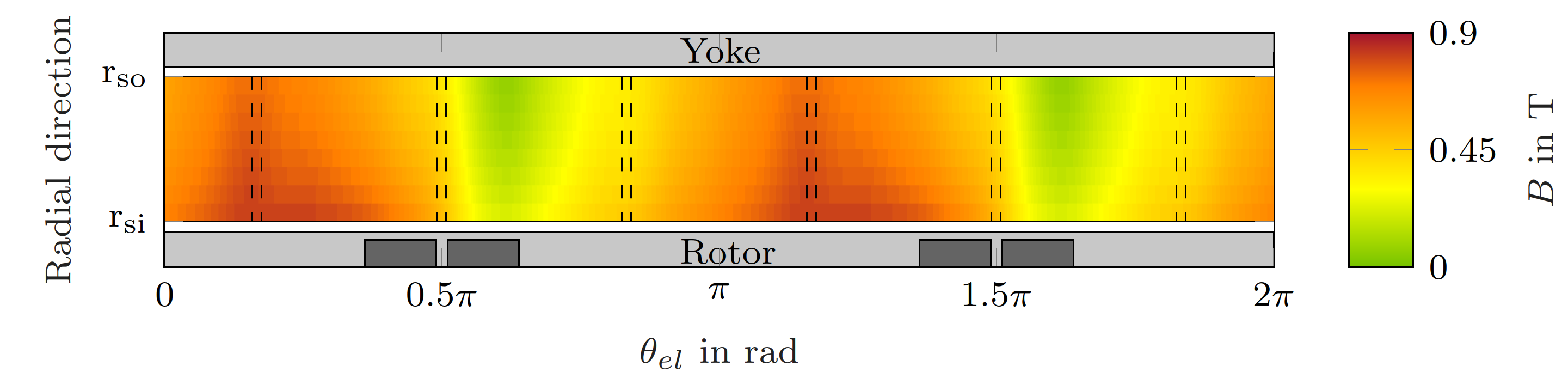}
		}
		\vskip 0.0\textwidth
		\subfloat[Normalized current\label{subfig:In_r_theta}]{
			\centering
			\def\svgwidth{0.33\textwidth}
			\includegraphics{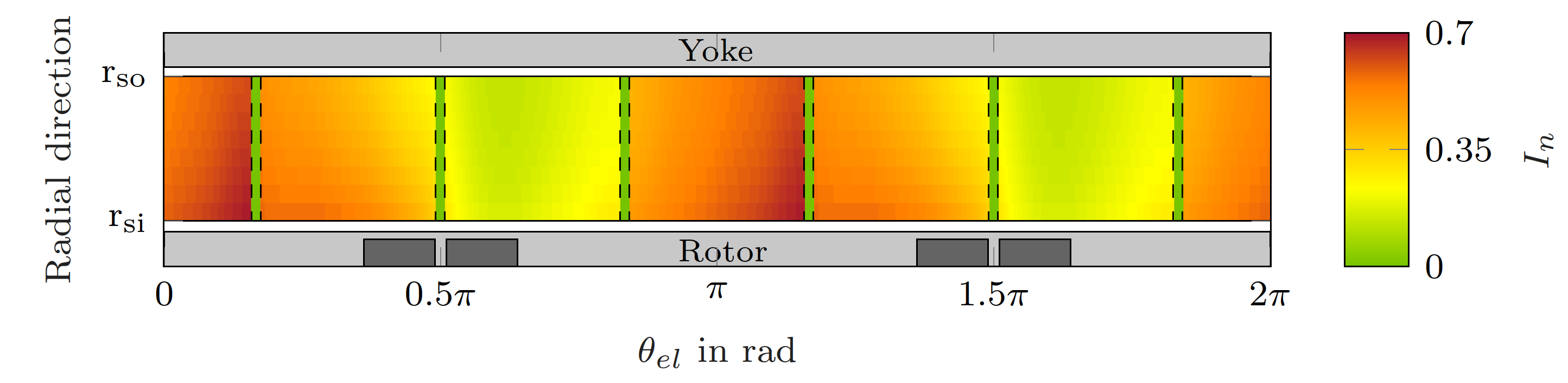}
		}
		\vskip 0.0\textwidth
		\subfloat[Total AC loss\label{subfig:Pv_r_theta}]{	
			\centering
			\def\svgwidth{0.33\textwidth}    
			\includegraphics{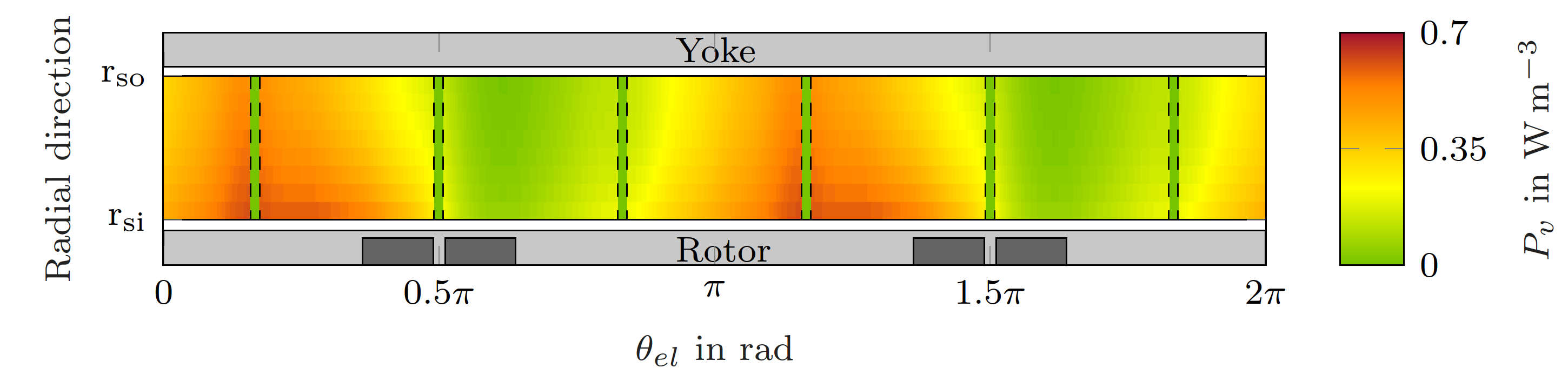}
		}
		\caption{Results for the fully SCM with the highest $PTW$ shown in Figure~\ref{subfig:PTW_Bwp_all_MgB2}. It is shown the stator current density $J_s$ \protect\subref{subfig:Jphases_theta} as a function of the electric angle $\mathrm{\theta_{el}}$ and the flux density $B$ \protect\subref{subfig:B_r_theta}, the normalized current $I_n$ \protect\subref{subfig:In_r_theta} and the total AC loss $P_{v}$ \protect\subref{subfig:Pv_r_theta} as a function of the radial direction $r$ and the electric angle $\mathrm{\theta_{el}}$. Yoke and rotor carrier are marked \legendsquare{fill=grey200} and rotor coils are marked \legendsquare{fill=grey100} as well as coil edges.}
		\label{fig:2DslotAClosses}
	\end{center}
\end{figure}

Additionally, Figure~\ref{subfig:PTW_Jwp_all_MgB2} and \ref{subfig:PTW_Jwp_all_Cu} show the torque-to-weight ratio $TTW$ depending current density $J_{wp}$ in the working point. The MgB\textsubscript{2} wire enables current densities up to 8 times higher compared to copper. The calculation of the current density includes the specific area required for cooling channels in both concepts. In each design $J_{wp}$ and $B_{wp}$ are linked via the electro-thermal behavior of the conductors, wherefore the current density plots in Figure~\ref{subfig:PTW_Jwp_all_MgB2} and \ref{subfig:PTW_Jwp_all_Cu} are inversely related to the flux density plots in Figure~\ref{subfig:PTW_Bwp_all_MgB2} and \ref{subfig:PTW_Bwp_all_Cu}, respectively.\\
\begin{figure}
	\hspace{-0.55cm}
	\subfloat[Fully SCM with the 5 highest PTWs
	\label{subfig:PTW_EFF_MgB2}]{
		\centering
		\def\svgwidth{0.33\textwidth}
		\includegraphics{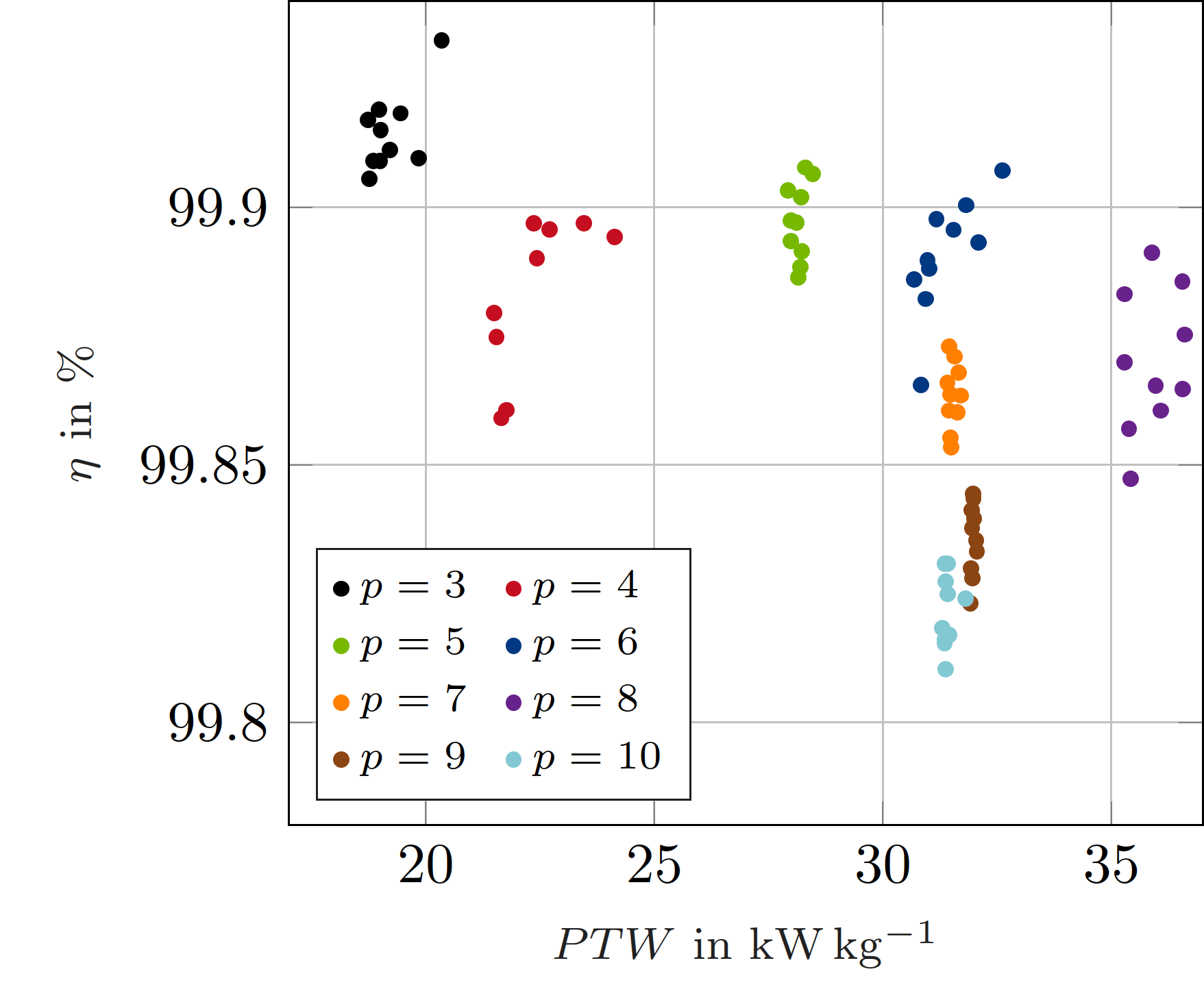}
	}
	\hspace{0.2cm}
	\subfloat[Partially SCM with the 5 highest PTWs
	\label{subfig:PTW_EFF_Cu}]{
		\centering
		\def\svgwidth{0.33\textwidth}
		\includegraphics{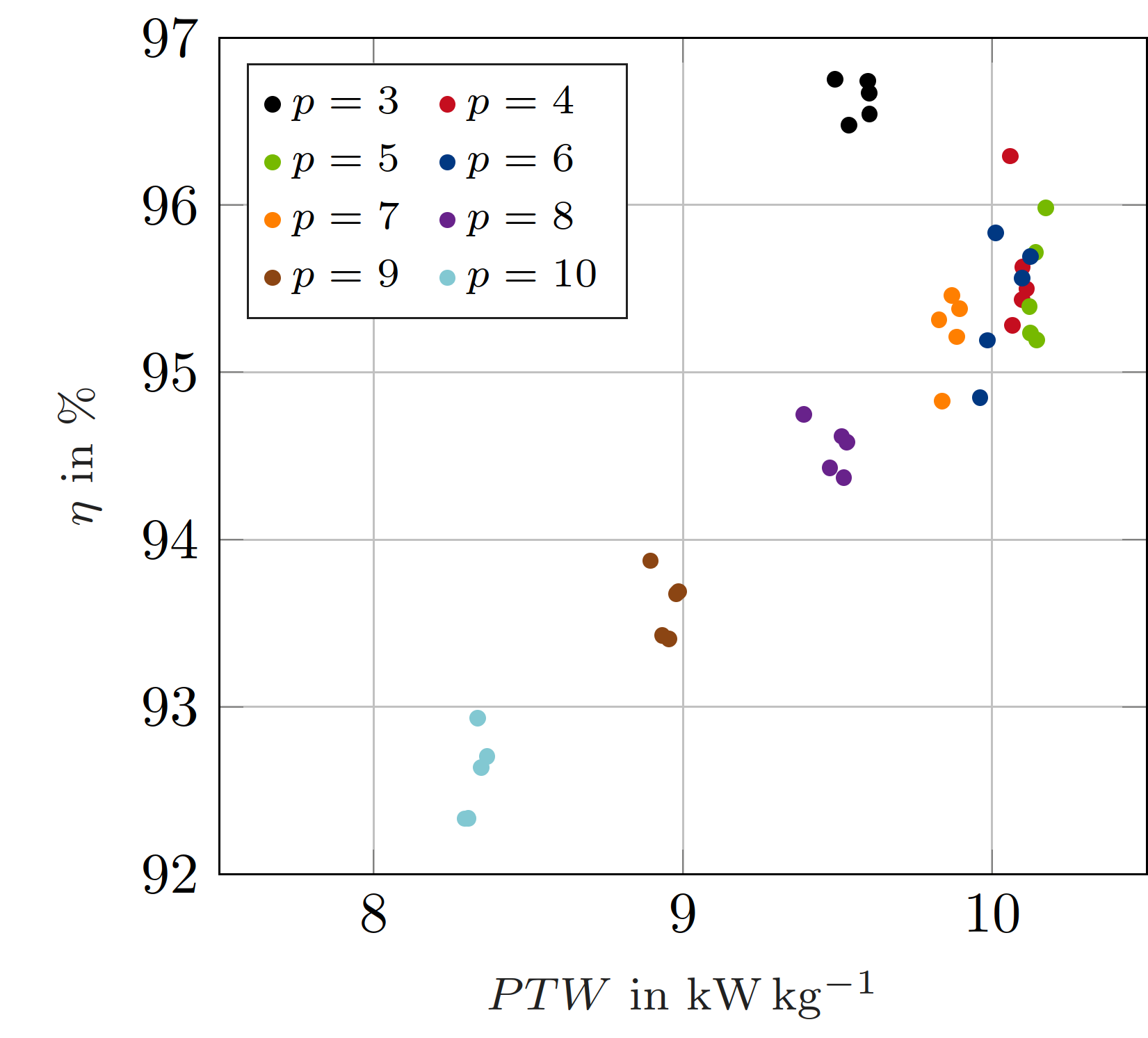}
	}
	\vskip 0.00\textwidth
	\subfloat[Fully SCM with the 5 highest PTWs
	\label{subfig:PTW_Pv_rotor_MgB2}]{	
		\centering
		\def\svgwidth{0.33\textwidth}         
		\includegraphics{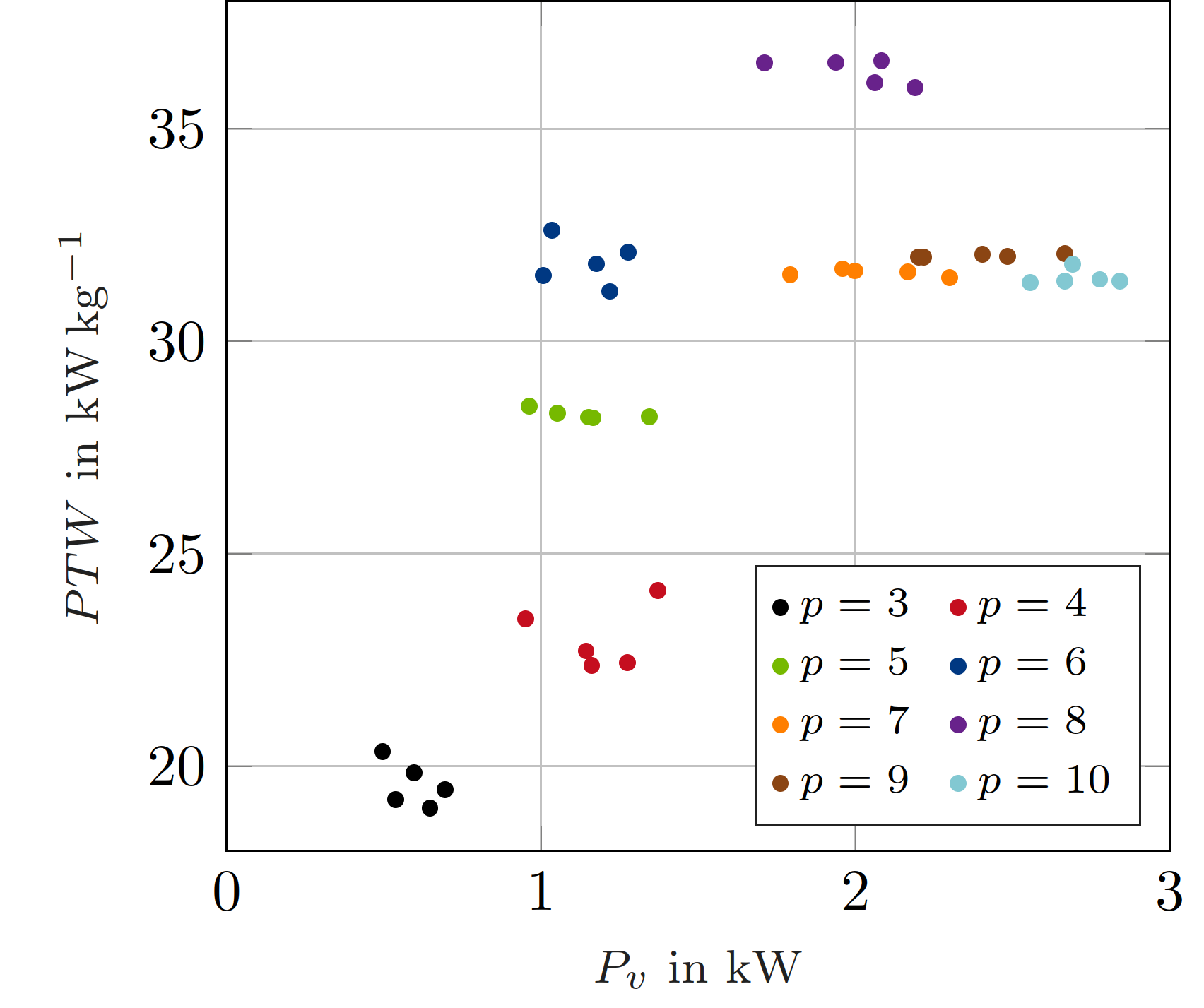}
	}
	\subfloat[Partially SCM with the 5 highest PTWs
	\label{subfig:PTW_Pv_rotor_Cu}]{	
		\centering
		\def\svgwidth{0.33\textwidth}           
		\includegraphics{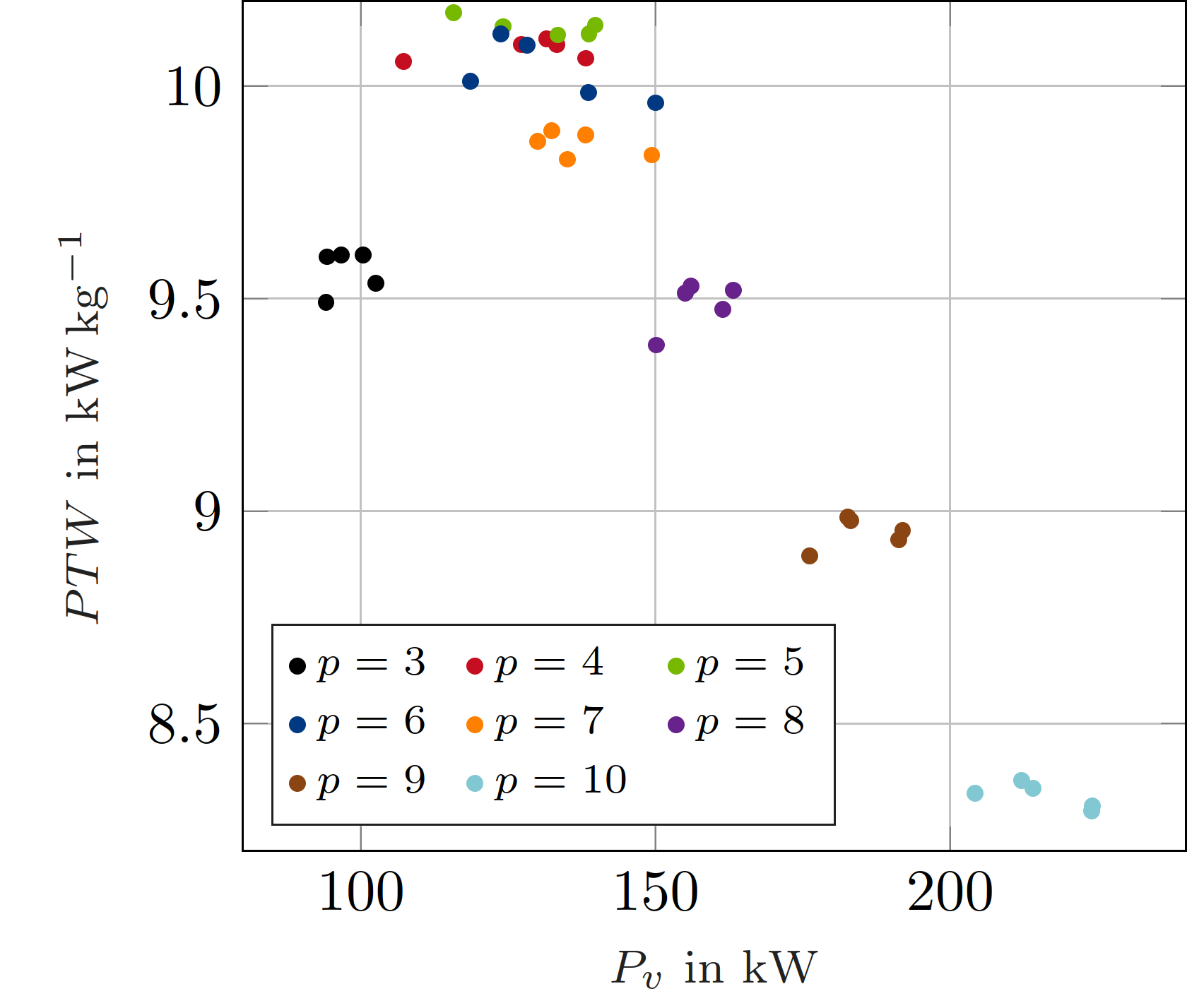}
	}
	\vskip 0.00\textwidth
	\subfloat[Fully SCM with the 100 highest PTWs
	\label{subfig:PTW_r_b_rotor}]{
		\centering
		\def\svgwidth{0.33\textwidth}    
		\includegraphics{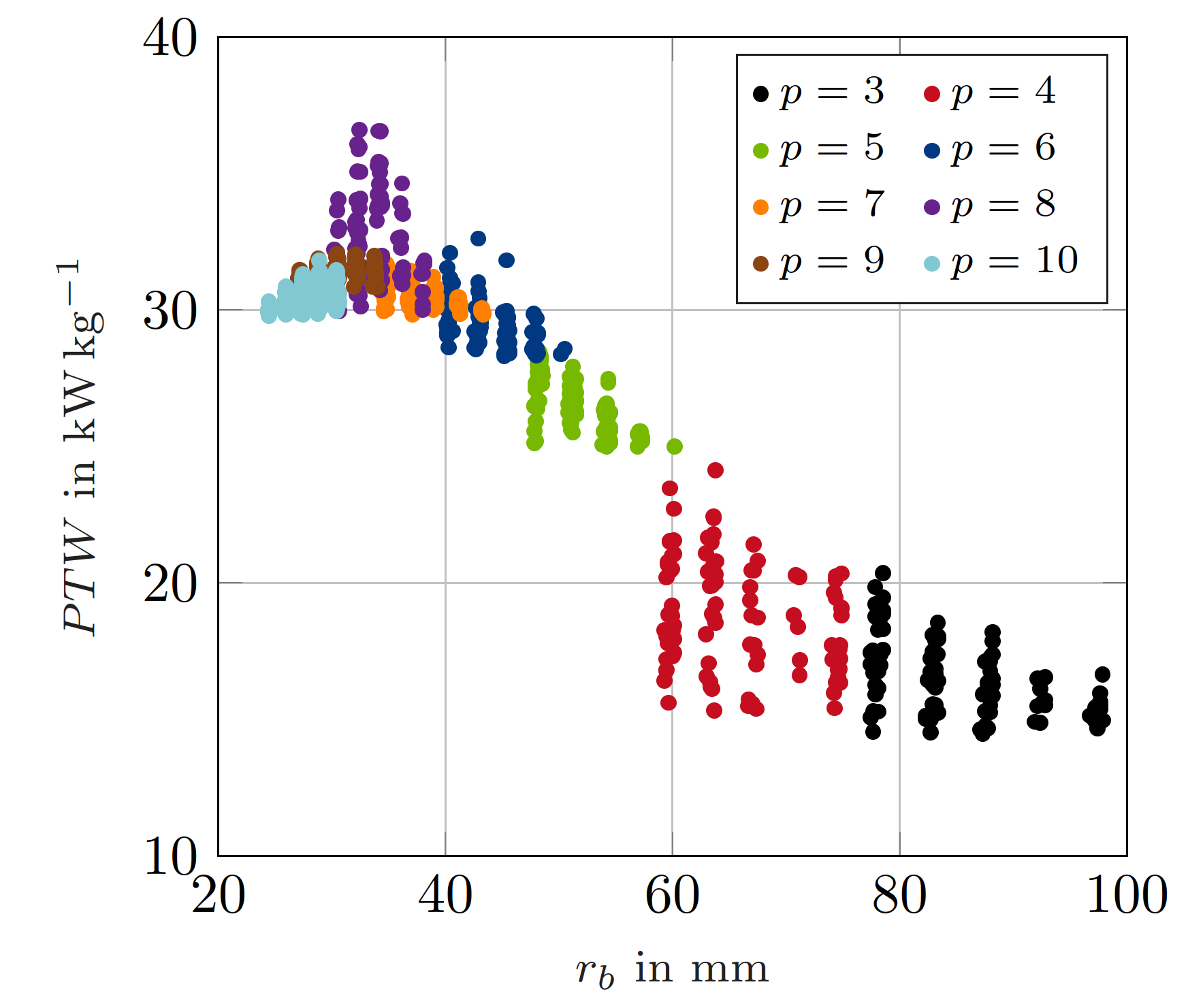}
	}
	\hspace{-0.2cm}
	\subfloat[Fully SCM with the 100 highest PTWs
	\label{subfig:PTW_dl_rotor}]{
		\centering
		\def\svgwidth{0.33\textwidth}
		\includegraphics{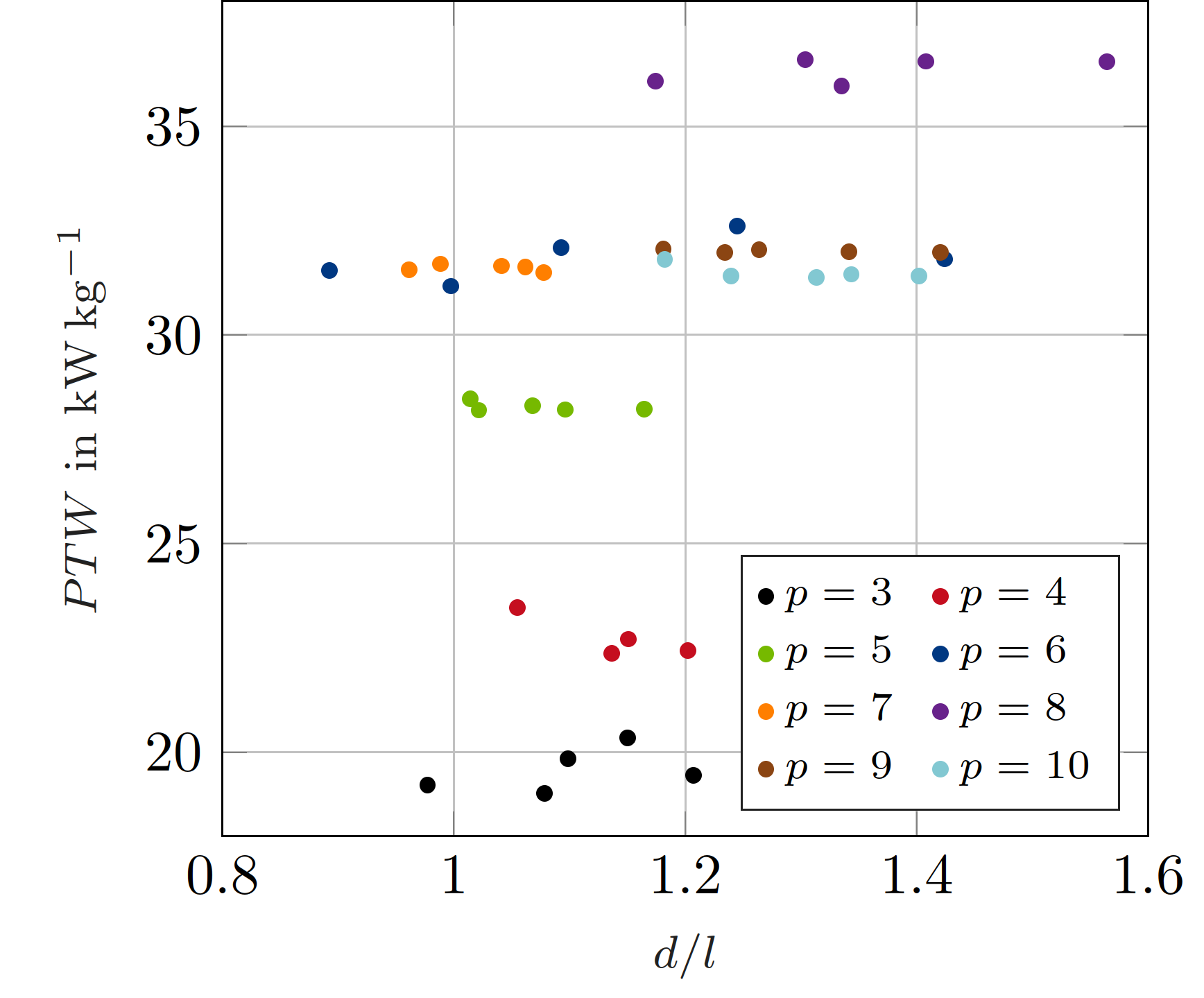}
	}
	\vskip 0.00\textwidth
	\caption{Power-to-weight ratio $PTW$ as a function of the efficiency $\eta$ in \protect\subref{subfig:PTW_EFF_MgB2}, \protect\subref{subfig:PTW_EFF_Cu}, the total stator AC loss $P_{v}$ in \protect\subref{subfig:PTW_Pv_rotor_MgB2}, \protect\subref{subfig:PTW_Pv_rotor_Cu}, the stator bending radius $r_b$ in \protect\subref{subfig:PTW_r_b_rotor} and the diameter-to-length aspect $d/l$ in \protect\subref{subfig:PTW_dl_rotor} for pole pair numbers $p$ between $3$ $(225\, \mathrm{Hz})$ and $10$ $(750\, \mathrm{Hz})$.}
	\label{fig:results_PTW_coil2}  
\end{figure}
For the five designs with the highest PTW at each pole pair number, the dependency between the efficiency $\eta$ and the $PTW$ is presented in Figure~\ref{subfig:PTW_EFF_MgB2} for the fully SCM and in Figure ~\ref{subfig:PTW_EFF_Cu} for the partially SCM. For fully SCM the range of efficiencies is between $99.82 \, \%$ and $99.94 \, \%$, thus roughly two orders of magnitude higher than for the  partially SCM with efficiencies of $91.8 \, \%$ to $97.1 \, \%$. This difference is linked to stator losses that approximately two orders of magnitude smaller for the superconducting stator. The stator losses are shown in Figure~\ref{subfig:PTW_Pv_rotor_MgB2} and \ref{subfig:PTW_Pv_rotor_Cu}. In the case of the copper stator approximately $90\%$ of the loss are ohmic losses which sets the large offset of the x-axis. 
Interestingly, in the partially SCM case, the efficiency and $PTW$ are not inversely proportional as it is typical for synchronous machines but rather directly proportional. This is due to the fact that higher pole pair numbers lead both to higher losses and consequently lower current densities, thus to lower $PTW$.\\
Figure~\ref{subfig:PTW_r_b_rotor} shows the bending radius of the stator coils $r_b$ as a function of $PTW$ for the 100 lightest machines of each number of pole pairs. If the number of pole pairs increases, the bending radius decreases and is smallest at $26 \, \mathrm{mm}$ for a machine with 10 pole pairs. The minimum bending radius depends on the superconducting wire design and leads to critical current degradation. A typical MgB\textsubscript{2} wire has a minimum bending radius of about $40 \, \mathrm{mm}$ \cite{Kovac.2016b}. However, this effect strongly depends on the conductor design and was not taken into account in the model.\\
The dependency of the machines diameter-to-length aspect ratio $d/l$ and the $PTW$ is presented in Figure~\ref{subfig:PTW_dl_rotor}. In contrast to conventional machines, the $PTW$ is mostly independent on the aspect ratio.\\
The machine parameters and results for the lightest fully and partially SCM machine are summarized in Table~\ref{tab:results_parameters_example} including the Esson coefficient $C_e$ and total machine length $l$.
\begin{table}[!h]
	\caption{\label{tab:results_parameters_example}Machine parameters and calculation results for the lightest fully and partially SC machine designs. The calculation time is referring to the calculation of all 4320 designs.}
	\begin{center}
		\begin{tabular}{llll}
			\hline
			Symbol 			& Unit 						& Fully SCM			&  Partially SCM\\
			\hline
			$PTW$			& $\mathrm{kW \, kg^{-1}}$	& $36.6$ 			& $10.2$ \\
			$TTW$			& $\mathrm{Nm \, kg^{-1}}$	& $77.7$			& $21.6$ \\
			$\eta$			& $\mathrm{\%}$				& $99.87$			& $96$ \\
			$p$				& $\mathrm{-}$				& $8$				& $5$ \\
			$f_{el}$		& $\mathrm{Hz}$				& $600$				& $375$	\\
			$r_{ri}$		& $\mathrm{m}$				& $0.15$			& $0.16$ \\
			$d_{s,c}$		& $\mathrm{mm}$				& $10$				& $20$ \\
			$d_{m}$			& $\mathrm{mm}$				& $12$				& $12$\\
			$\alpha_{2}$	& $\mathrm{^\circ}$			& $68$				& $55$ \\
			$\alpha_{3}$	& $\mathrm{^\circ}$			& $1.91$			& $1.79$ \\
			$d_{mag}$		& $\mathrm{mm}$				& $8.4$				& $11.7$ \\
			$m_a$			& $\mathrm{kg}$				& $36.8$			& $160.4$ \\
			$m_p$			& $\mathrm{kg}$				& $45.1$			& $134.5$ \\
			$B_{wp}$		& $\mathrm{T}$				& $0.64$			& $1.08$\\
			$J_{wp}$		& $\mathrm{A \, mm^{-2}}$	& $177.2$			& $21.6$ \\
			$P_{v,y}$		& $\mathrm{kW}$				& $1.66$			& $4.76$ \\
			$P_{v,s}$		& $\mathrm{kW}$				& $2.08$			& $115.74$ \\
			$\dot{M}_{co}$	& $\mathrm{g/s}$			& $0.463$			& $19426$ \\
			$T_{in}$		& $\mathrm{K}$				& $20$				& $353$ \\
			$r_{b,s}$		& $\mathrm{mm}$				& $32.4$			& $56.7$ \\
			$l_{eff}$		& $\mathrm{mm}$				& $225.3$			& $407.5$ \\
			$l$				& $\mathrm{mm}$				& $292.4$			& $522.0$ \\
			$C_e$		& $\mathrm{kW \, min \, m^{-3}}$				& $26.1$				& $12.4$ \\
			$T_{\mathrm{cal}}$ & $\mathrm{h}$			& $21$			& $11$ \\
			\hline
		\end{tabular}
	\end{center}
\end{table}
The calculation time $T_{\mathrm{cal}}$ of the fully SCM is longer than that of the partially SCM due to the non-linearity of the superconductor. Both designs show a higher $PTW$ with the single pancake coil compared to the double pancake coil. For partially SCM this is due to the higher necessary thickness of the sleeve which enlarges the magnetic airgap. For fully SCM the required excitation fields can be handled by a single layer coil. \\
Figure~\ref{fig:2DslotAClosses} shows the results of the two-dimensional calculation of the AC loss of the stator coils, described in Section~\ref{electro_thermal_model}, step by step for the fully SCM with machine parameters as listed in Table~\ref{tab:results_parameters_example}.

The current density distribution of a 3-phase winding system is illustrated in Figure~\ref{subfig:Jphases_theta} and the spatial distribution of the normalized current is shown in Figure~\ref{subfig:In_r_theta}. The current-free areas between the coils are visible. The normalized current distribution does not exceed its maximum value of $I_{n,max} = \mathrm{0.7}$ which is a requirement. A loss hot spot is detected at the inner stator radius $r_{si}$ in Figure~\ref{subfig:Pv_r_theta}. This hotspot does not necessarily have to occur in the coil which currently carries the highest current, because the loss is still strongly dependent on the magnetic field, shown in Figure~\ref{subfig:B_r_theta}.

\section{Conclusion}

We presented an analytical model for the design of superconducting radial flux machines. The electromagnetic design is considered by a two-dimensional approach which takes into account the mechanic, as well as the thermodynamic limits of the parts in the air gap. Furthermore, the AC loss and the consumption of the coolant in the stator winding system is calculated. We find that the model can provide results very quickly and is therefore useful for large parameter scans in a preliminary machine design.\\
The influence of the material and geometry parameters on the performance of a fully and partial SCM was investigated exemplary based on requirements derived from the N3-X. We come to the conclusion that for fully SCM maximum power-to-weight ratios of $36.6 \, \mathrm{kW \, kg^{-1}}$ at an efficiency of $99.88\%$ while partially SCM maximum power-to-weight ratios of $10.2 \, \mathrm{kW \, kg^{-1}}$ at an efficiency of $96\%$ could be achievable, i.e. the fully superconducting machine is roughly 3.5 times lighter. This points out the high potential for fully superconducting machines. To compare the masses of the different topologies fairly, the penalty masses of the required cooling systems need to be taken into account. To calculate their required size the efficiency and cooling requirements of each machine needs to be computed for every different power requirement along with the mission profile of the aircraft. Even if our model takes into account the mass of most passive components, further detailing which includes bearings, shaft, instrumentation, and high voltage connectors is desireable. This will reduce the $PTW$ a bit, but we assume that $PTW$ values larger than $30 \, kW \, kg^{-1}$ are realistic. Therefore, due to our coupled electric and thermal modeling, our approach provides to accomplish this in various future studies of hybrid-electric aircraft. Our analysis concludes that the required $PTW$ of $12.7 \, \mathrm{kW \, kg^{-1}}$ for the N3-X concept aircraft is only feasible with the fully superconducting machines.\\
While partially superconducting machines reached a TTW value of up to $21.6 \, \mathrm{Nm \, kg^{-1}}$ which is comparable to non-superconducting machines, fully superconducting machines showed results beyond $75 \, \mathrm{Nm \, kg^{-1}}$.  
In combination with decreasing losses in the stator of fully superconducting machines when lowering the electric frequency, we suggest that this machine type might be particularly interesting to be studied for even lower speed direct drive applications, such as propellers or large fans \cite{Cameretti.2018}.\\ 
We find also that the best power-to-weight ratios come with designs with comparatively low magnetic flux densities in the airgap in the range from 0.55\,T to 0.9\,T - values that could also be achieved with NdFeB magnets. This can be attributed to the high sensitivity of the current-carrying capacity of the MgB\textsubscript{2} wire to external fields and frequencies. Consequently, we can conclude that material development should focus on improving the AC loss, current-carrying capacity and bending properties of superconducting wires rather than achieved extreme magnetic fields in the rotor.\\
Further, an investigation on partially superconducting machines with a superconducting stator and a rotor with Halbach-NdFeB magnets appears interesting. The mechanical effort in a rotating superconducting rotor is eliminated in this approach. However, Halbach magnets may have the potential to achieve high power-to-weight ratios due to the small magnetic air gap in such a machine. 

\section{Acknowledgments}

The authors acknowledge the financial support by the Federal Ministry for Economic Affairs and Energy of Germany in the framework of LuFoV-2 (project number 20Y1516C). We thank Stefan Moldenhauer, Stefan Biser, Joern Grundmann and Mabroor Ahmed for helpful discussions.

\FloatBarrier
\section*{References}
\bibliography{arXiv}

\end{document}